# Signatures of a high-temperature collective electronic phase with superconductivity-like characteristics and a giant pressure effect in networks of boron-doped ultrathin carbon nanotubes

Y. Wang[1,†], T. H. Koo[1,†], R. Huang[1,†],Y. H. Ng[1,†], T. T. Lortz[1,†], T. Zhang[1,†], W. M. Chan[1], Y. Hou[1], J. Pan[1], S. Krämer[2], A. Demuer[2], R. Lortz[2,*,♠], N. Wang[1,*,♣], and P. Sheng[1,3,*,♥]

[1]Department of Physics, The Hong Kong University of Science and Technology, Kowloon, Hong Kong, China

[2]Université Grenoble Alpes, Université Toulouse, INSA Toulouse, CNRS, LNCMI, EMFL, Grenoble, France.

[3]Clare Hall College, Cambridge University, Cambridge, The United Kingdom

[†]These authors contributed equally to this work and share first authorship

[*]These authors contributed equally to this work and share last authorship

[♠]Contact author: rolf.lortz@lncmi.cnrs.fr

[♣]Contact author: phwang@ust.hk

[♥]Contact author: sheng@ust.hk

**ABSTRACT**

We present data consistent with a high-temperature collective electronic phase with superconductivity-like characteristics in three dimensional networks of boron doped, ultrathin carbon nanotubes (CNTs) grown inside the ~5 Å channels of ZSM 5 zeolite. Confinement stabilizes (2,1) CNTs that are otherwise dynamically unstable, while boron doping shifts the Fermi level toward a van Hove singularity, as supported by ab-initio calculations. The resulting CNT network exhibits multiple, mutually consistent signatures of an electronic condensate with the typical characteristics of a high temperature superconductor at ambient pressure. DC magnetization and AC susceptibility measurements reveal the onset of a Meissner response between 220 and 250 K, with compacted samples achieving up to 93% of full diamagnetic screening. Electrical transport shows a sharp resistive transition with extrapolated $T_c \approx 239$ K and vanishing resistance in optimized samples. Specific heat measurements display a reproducible anomaly at 233–236 K that broadens under magnetic field, consistent with strong fluctuations. Point contact spectroscopy identifies three energy gaps, including a leading gap of ~30 meV whose temperature dependence follows BCS expectations for $T_c \approx$ 224 K, and exhibits particle–hole symmetry and Andreev reflection. Remarkably, applying pressures below 0.1 kbar enhances $T_c$ by nearly 100 K and modulates the room temperature resistance by more than three orders of magnitude, suggesting a pressure driven 1D to 3D crossover in the CNT network. These results identify boron doped ultrathin CNT networks as a promising carbon-based platform for near ambient temperature superconductivity and reveal an unusually large pressure sensitivity with potential technological relevance.

## I. INTRODUCTION

The pursuit of superconductivity at elevated temperatures has driven major advances in condensed-matter physics for more than a century [1]. Following the discovery of superconductivity in 1911 [2], conventional phonon-mediated superconductors were long believed to be constrained by the McMillan limit [3], which places an upper bound of order 40 K on the critical temperature $T_c$ within the Bardeen–Cooper–Schrieffer (BCS) framework [4]. The emergence of cuprate superconductors in 1986, with critical temperatures exceeding 130 K at ambient pressure, demonstrated that unconventional pairing mechanisms can circumvent this limit [5-7]. Subsequent discoveries in iron-based superconductors [8], nickelates [9,10], and high-pressure hydrides have further expanded the landscape of materials exhibiting superconducting-like behavior at high temperatures [11-14]. Yet, despite these advances, the realization of robust superconductivity near ambient conditions remains an outstanding challenge.

Carbon nanotubes are cylindrical nanostructures made of carbon atoms, renowned for their extraordinary strength, electrical conductivity, and potential to revolutionize fields from electronics to medicine [15-17]. Whether thin carbon nanotubes can exhibit high-$Tc$ superconductivity [18] has long been a tantalizing question, as carbon-based materials present a distinctive pathway toward high-temperature electronic phenomena. Their light atomic mass, diverse bonding configurations, and tunable low-dimensional electronic structures have led to reports of superconductivity in doped fullerenes [19], carbon nanotube (CNT) bundles [20], and twisted bilayer graphene [21]. In particular, ultrathin CNTs confined within zeolite channels have been proposed as a platform where strong one-dimensional (1D) electronic correlations, enhanced density of states near van Hove singularities, and enhanced electron-phonon coupling [22] may combine to promote pairing at elevated temperatures [23-28]. However, experimental signatures in such systems have remained incomplete or difficult to reproduce, leaving open the question of whether confinement-engineered CNT networks can host a phase-coherent electronic state at high temperature.

In this work, we investigate boron-doped ultrathin CNTs grown within the ~5 Å channels of ZSM-5 zeolite [28], forming a three-dimensional (3D) network of mechanically coupled 1D conductors. Ab-initio calculations indicate that the (2,1) CNTs stabilized by confinement possess soft phonon modes and a high density of states near the Fermi level when appropriately doped. The resulting CNT@ZSM-5 composite therefore provides a unique setting in which 1D electronic structure, van Hove singularity tuning, and inter-tube coupling may combine to generate collective electronic behavior.

We present a set of experimental observations—spanning electrical transport, AC and DC magnetic susceptibility, specific heat, and point-contact spectroscopy—that consistently reveal the onset of a phase-coherent state between 220 and 250 K at ambient pressure. Each probe exhibits features commonly associated with superconductivity, including a sharp resistive drop, diamagnetic screening, a thermodynamic anomaly, and particle–hole symmetric gap-like spectra with Andreev reflection. While none of these signatures alone is conclusive, their concurrence at a common temperature scale across multiple batches and measurement techniques suggests the emergence of a collective electronic phase with superconducting-like characteristics.

An additional and unexpected finding is the pronounced sensitivity of this state to modest applied pressure. Pressures below 0.1 kbar, applied through mechanical compaction of the powder, shift the onset of the resistive transition upward by nearly 100 K and strongly modulate the room-temperature resistance. Such a large pressure response is consistent with the mechanically fragile nature of the CNT network and the proximity of the system to a 1D–3D crossover in inter-tube coupling. Regardless of its microscopic origin, this tunability represents a technologically relevant property of the CNT@ZSM-5 composite.

Taken together, these results identify boron-doped ultrathin CNT networks as a promising platform for exploring high-temperature phase-coherent electronic states in low-dimensional carbon systems. While further work is required to establish the microscopic pairing mechanism and to fully confirm the superconducting nature of the observed phenomena, the convergence of multiple experimental signatures motivates deeper investigation of this material class and its unusual pressure response.

In what follows, we first describe the sample fabrication and characterization procedures in Section II, then outline the sample structure and its physical implications in Section III. The experimental results are presented in Section IV, with discussion and concluding remarks provided in Sections V and VI, respectively. Details of the experimental measurement methods and selected aspects of the ab-initio calculations are given in Section VII to avoid digressing from the main line of exposition.

## II. SAMPLE FABRICATION AND CHARACTERIZATION

Prior to synthesis, commercial ZSM-5 crystals [29] were calcined at 550°C for 6 hours in air to remove organic residues and adsorbed water, ensuring clean pore channels. Boron-doped CNT networks were synthesized in ZSM-5 via CVD, as described in the following. The calcined zeolite was loaded into a quartz boat, degassed at 200°C under vacuum ($10^{-3}$ mbar) for 2 hours, and then heated to 910°C at 24°C/min. At 910°C, the in situ-generated diborane and hydrogen gas ($B_2H_6$ and $H_2$, 62 cm³, 1.4 atm) and high-purity methane (99.99% and 200 sccm at 6 atm) were sequentially released into the CVD chamber (200 cm$^3$) and co-flowed through the zeolite bed (setup shown in Fig. 1). Diborane was synthesized via the reaction of small amounts of iodine and sodium borohydride ($I_2 + 2NaBH_4 \rightarrow 2NaI + B_2H_6 + H_2$) at 180°C for 2 hours in a sealed vessel; the synthesized diborane was then used as the boron source. The total system pressure was maintained at 6 atm for 5 hours using a MKS651C® controller. Owing to the imperfect sealing of the CVD chamber, some residual methane flow was necessary to maintain a constant pressure. This residual methane flow inevitably caused a gradual decrease in the boron concentration, necessitating the reintroduction of diborane after 2 and 4 hours to maintain the uniformity of the doped material over time. After 5 hours, the methane source was closed, and the chamber was evacuated. Controlled cooling to 750°C (0.6°C/min) was carried out over 4.5 hours to facilitate surface carbon sublimation under the partial vacuum condition, and sample annealing. The cooling/annealing time of 4.5 hours was selected to retain a thin layer of carbon on each grain surface. The oven was subsequently turned off to allow natural cooling of the material to room temperature.

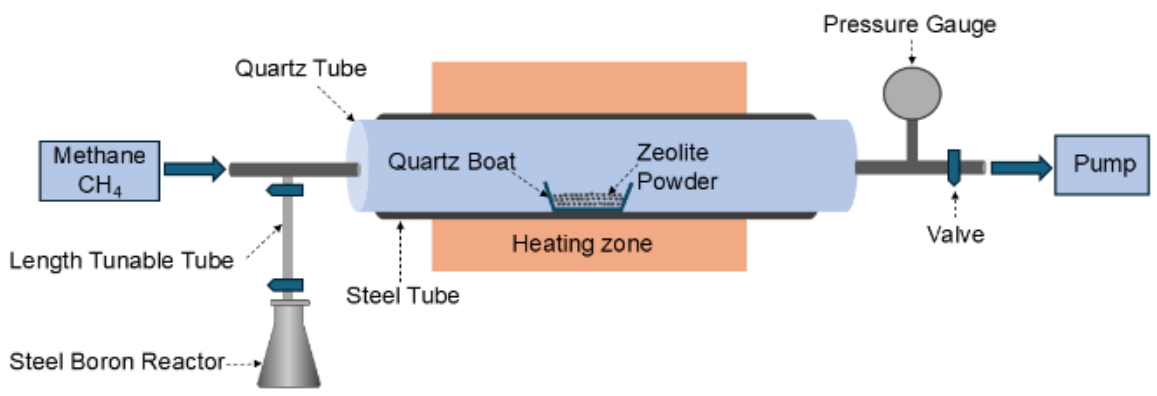

FIG 1. Chemical vapor deposition and boron-doping setup for the fabrication of doped carbon nanotube networks in ZSM-5, including a source of methane gas, a steel boron reactor, a quartz tube within an oven in which a quartz boat with the zeolite powder is placed, a pressure gauge and a pump.

The Raman spectroscopy data (Jobin Yvon T64000®, 514 nm laser excitation) reveal two prominent Raman peaks: the $G^+$ band (1586 cm$^{-1}$) and the $G^-$ band (1345 cm$^{-1}$) (Fig. 2a). Whereas the $G^+$ mode is indicative of the relative C–C vibration of the $\sigma$ bond along the axial direction, the $G^-$ mode configuration is indicative of the strong $\sigma\pi$ bond mixing in the (2,1) CNT and hence has a value close to the G band frequency of 1332 cm$^{-1}$ of diamond. In the present case, the $G^+$ band matches reasonably well with the ab-initio calculated value of 1540 cm$^{-1}$ for the (3,0) CNT. The $G^-$ band matches well with the ab initio calculated value of 1363 cm$^{-1}$ for the (2,1) CNTs.

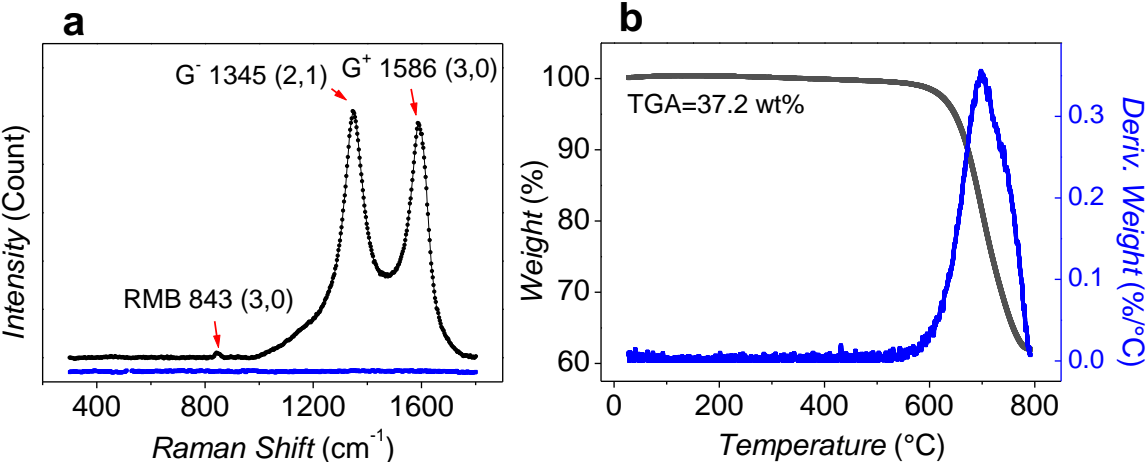

FIG 2. Characterization of boron-doped carbon networks in ZSM-5. **a**, Raman spectrum showing two sharp G band peaks, the $G^+$ peak for the (3,0) CNTs at 1586 $cm^{-1}$ and the $G^-$ peak for the (2,1) CNTs at 1345 $cm^{-1}$. **b**, Thermogravimetric analysis (TGA) diagram showing the peak structure in the differential (blue) curve. The TGA loss curve (black) shows a total of 37.2 wt% carbon, corresponding to a slightly over 100% pore occupation ratio with some intended carbon surface covering.

The calculated Raman radial breathing mode (RBM) frequency is 813 $cm^{-1}$ for the (3,0) CNTs, which agrees reasonably well with the small Raman peak at 843 $cm^{-1}$. However, the calculated Raman frequency of the (2,1) CNTs is only 699 $cm^{-1}$ owing to their instability (Suppl. Fig. S1) [30]. The stability of the (2,1) CNTs is maintained by the confinement pressure of the channel wall. Because the confinement pressure is unknown, the RBM frequency of the (2,1) CNTs cannot be determined.

The results of the thermogravimetric analysis (TGA) (Q5000(TA)) are shown in Fig. 2b, which revealed that the carbon content to be 37.2 wt%. Since the 100% pore-filling wt% is ~29 wt%, samples processed under 4.5 hours of annealing time have a thin surface carbon layer. In earlier samples, the duration for annealing was set at 5 hours, which led to samples with a TGA value of 27.9 wt%, corresponding to a 94% pore-filling ratio. Hence by adjusting the parameter values like the cooling duration for annealing from 910 to 750 C, the carbon content can be adjusted, to more or less than 29 wt%. A thin surface carbon layer on the grain can facilitate electrical grain contact.

Energy-dispersive X-ray spectroscopy was used to determine the boron doping level, yielding a carbon-to-boron (C:B) ratio between 10:1 and 9:1. This doping shifts the Fermi level close to a van Hove singularity, which is essential to obtain a large density of states at the Fermi level, a key ingredient for a high superconducting transition temperature (Suppl. Fig. S1) [30].

## III. DESCRIPTION OF THE MATERIAL

ZSM-5 zeolite, a commercially-available microporous aluminosilicate characterized by a three-dimensional interconnected pore network, serves as a template for the synthesis of boron-doped single-walled CNTs via chemical vapor deposition (CVD). The pore system consists of two types of channels: straight channels with diameters of 5.4 × 5.6 Å oriented along the b-axis and slightly undulating channels with an elliptical cross section of 5.1 × 5.5 Å along the a-axis (Fig. 3a). Typical ZSM-5 crystallites are 3–4 µm in size (Fig. 3b). The channels form a three-dimensional network with a lattice constant of approximately 1 nm. This sub-nanometer confinement restricts nanotube growth to diameters less than 0.3 nm. Furthermore, the distinct geometries of the channels along the a- and b-axes determine the type of ultrathin CNTs formed: (3, 0) CNTs form in the b-axis channels, and (2, 1) CNTs form in the a-axis channels, the latter arising from the slightly elliptical cross section with a reduced minor axis. Importantly, the perpendicular channels do not intersect directly (Fig. 3c–d); instead, they are offset by a small vertical separation of ~1.5 Å at their closest approach (Fig. 3e). At such a small separation, some bonding may occur, and these regions represent mechanically weak points in the system; these points may account for the large pressure dependence of $T_c$ arising from 1D to 3D crossover.

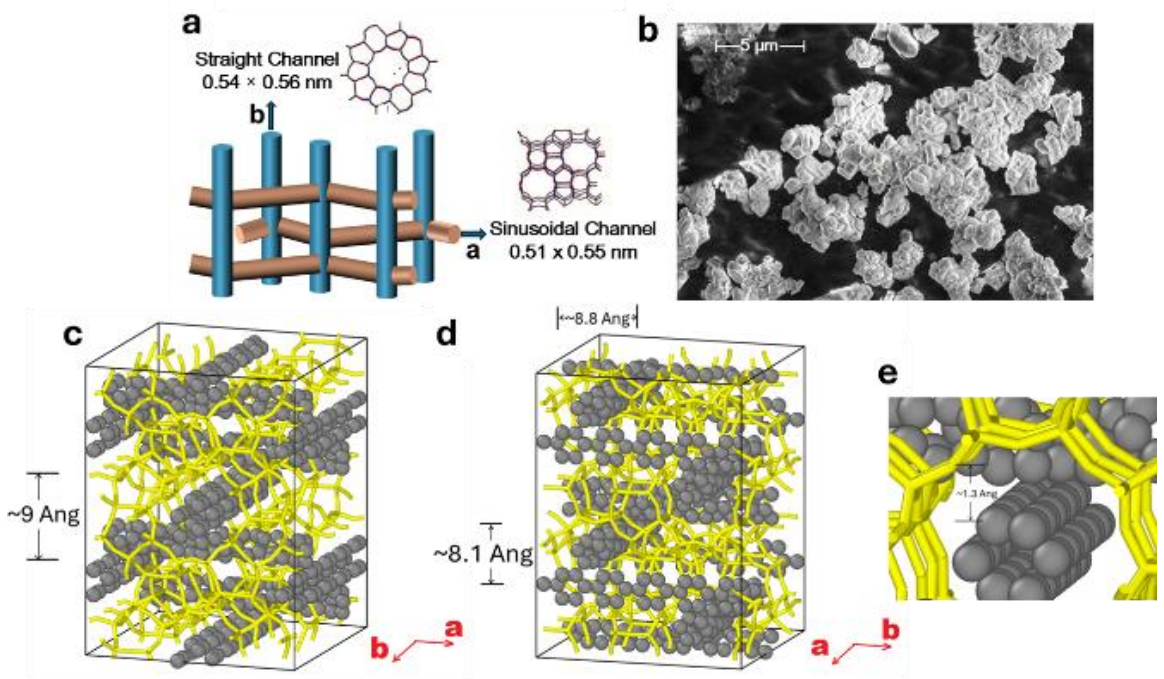

FIG 3. Crystalline structure of CNT@ZSM-5. **a**, Schematic of the pore structure of ZSM-5. **b**, Scanning Electron Microscopy (SEM) image of ZSM-5 zeolite crystallites with a typical size of 3– to 4 μm, which serve as a template for growing boron-doped networks of carbon nanotubes inside their pores to form CNT@ZSM-5. **c**, **d**, Illustration of the crystalline structure of the ZSM-5 zeolite, depicted via a skeletal framework (yellow) for visual clarity, with its pores filled with a 3D network of (2,1) CNTs (gray) in the channels along the a axis and (3,0) CNTs in the channels along the b axis. **e**, Detail of the region where the perpendicular nanotubes are separated by a small spatial gap of only 1.3 Å.

Ab-initio calculations were performed to derive the phonon spectra and electronic band structure. It was found that the phonon spectrum of the (2, 1) CNT contains imaginary frequencies across nearly all the wavenumbers, which is the typical signature of intrinsic instability and necessitating confinement by the zeolite channel walls to achieve structural stability (Suppl. Fig. S1) [30].

The filamentary structure exhibits several important physical property implications for a superconducting material. First, its very weak magnetic susceptibility suggests an exceptionally large effective magnetic penetration depth in samples composed of filamentary segments. This can lead to a weak dependence of observable superconducting properties—such as electrical resistance—on the applied magnetic field. More on this point is given in the discussion section.

Second, there is a transition from one-dimensional (1D) superconductivity, which does not exhibit zero resistance at finite temperatures due to fluctuations, to the emergence of global coherence within a three-dimensional (3D) network of filamentary structures hosting an electronic condensate. This 1D-to-3D crossover represents a central novelty of the present work.

## IV. RESULTS

### A. Search for the Meissner effect

Magnetization and susceptibility measurements are used to detect the diamagnetic Meissner effect—a hallmark of superconductivity. The DC magnetization data in normalized units of $4\pi\chi_{dc}$ measured for loose powder samples under zero-field cooled (ZFC) and field cooled (FC) conditions in magnetic fields of 200 Oe on the example of batch #67 were obtained (Fig. 4; other batches shown in Suppl. Figs. S2–S6 [30]). To enhance clarity, the weak linear background was subtracted, and the data were fitted at the highest temperature of 300 K (Suppl. Figs. S2–S4) [30]. Below an onset temperature that varies between 220 K and 250 K among the different batches (here, 250 K for batch #67), a negative contribution similar to the superconducting Meissner effect emerges under ZFC conditions, with a separation observed between the ZFC and FC branches, characteristic for flux pinning in a superconducting state (Fig. 4a). Data under different fields are shown in Suppl. Figs. S5 & S6 [30]. The onset of a weak diamagnetic signal is observed in the ZFC branch below 250 K, with the onset varying between 220 and 250 K—caused by small variations in growth conditions of the different sample batches—gives a first indication of a possible high-temperature electronic condensate in CNT@ZSM-5. The weakness of the signal indicates Meissner screening on the order of 0.01% perfect diamagnetism, which is partially due to the fine powder form, with crystallites much smaller than the London penetration depth that prevents the flow of macroscopic Meissner currents, as well as because of the filamentary nature of the CNT network.

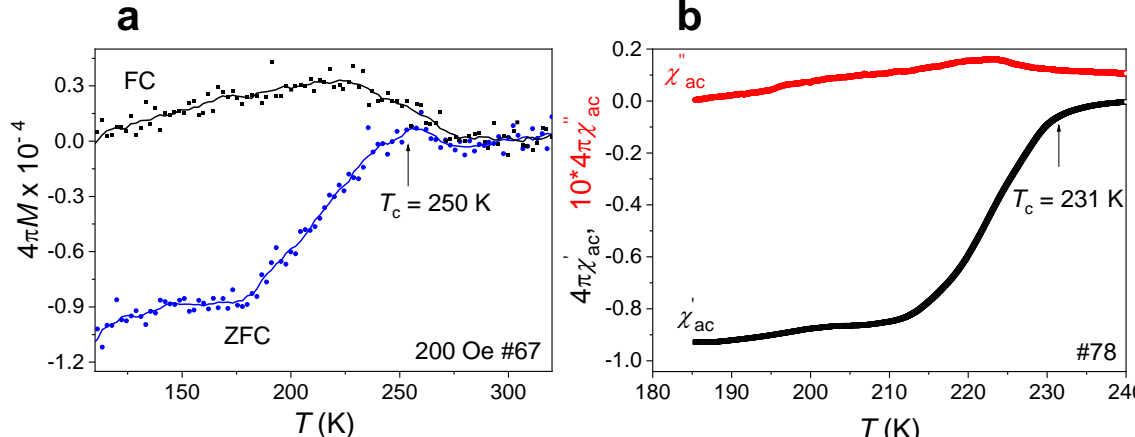

FIG 4. Signatures of a Meissner effect in DC magnetization and AC susceptibility of CNT@ZSM-5. **a** Example of DC susceptibility data in units of $4\pi\chi_{dc}$ (batch #67) measured under ZFC and FC conditions in a 200-Oe field. A weak linear background fitted above ~250 K was subtracted for clarity. The symbols represent the measured data, while the solid lines were generated by adjacent averaging over 7 points to increase clarity. **b** AC susceptibility at a frequency of ~1 kHz in units of $4\pi\chi$ measured with compacted powder samples with optimized surface treatment of the grains, showing a relatively sharp jump and peak in the real and imaginary parts χ' and χ'', respectively, below ~231 K for this sample batch (#78) with an absolute value reaching 93% of the perfect Meissner screening, calibrated with a $YBa_2Cu_3O_{6.92}$ reference sample.

To induce macroscopic Meissner currents and stronger Meissner screening to obtain a more definite proof of a possible superconducting origin of the observed diamagnetism, we attempted to compress the powder into pellets. However, the grains did not adhere together, the resulting pellets were very loose, and no significant enhancement of the presumed Meissner effect was observed. We therefore used an ac susceptibility setup (see section VII for more details) in which the powder was compacted by two screws within a polyamide block. This configuration allowed us to measure the AC susceptibility of a macroscopic sample in which the densely compacted powder allows for macroscopic screening currents. The magnitude of the signal was calibrated with a $YBa_2Cu_3O_{6.92}$ reference sample with the same dimensions and a signal value of -1 in the superconducting state in normalized units of $4\pi\chi_{ac}$, representing perfect Meissner screening (verified for $YBa_2Cu_3O_{6.92}$ with a SQUID magnetometer). In addition, we optimized the CNT@ZSM-5 surface by intentionally leaving a thin carbon layer on the crystallites, which greatly reduced the grain boundary resistance in such compacted powder samples. A step-like increase in the real component of the ac susceptibility signal $\chi_{ac}'$ and a peak in the imaginary component $\chi_{ac}''$ were observed below 230 K (Fig. 4b), both of which are characteristic signatures of superconductivity. For this compact macroscopic sample, 93% of the total Meissner screening was reached when an excitation frequency of 1 kHz was used and the sample was well compressed prior to the experiment. At present, the only known mechanism capable of producing such a strong Meissner effect is superconductivity.

The substantial discrepancy between the magnetic susceptibility values measured by SQUID and ACS can be attributed to the exceptionally large effective magnetic penetration depth associated with sample's intrinsic filamentary structure. This penetration depth may significantly exceed the effective thickness of the sample in the SQUID configuration (a powdery layer), while remaining smaller than the characteristic dimension in the ACS measurement, which exceeds 1 mm. This point is further addressed in the discussion Section V.

## B. Electrical resistivity and zero resistance

Electrical resistance measurements were carried out on ~1 mg of CNT@ZSM-5 powder compressed between either two or four ~1-mm-diameter screws serving as electrical contacts within a polyamide block (see section VII for more details). The latter configuration inadvertently enabled exploration of the pronounced pressure dependence of the sample as a byproduct, with the potential to increase $T_c$ above an ambient temperature, as discussed in the following section C. The two-probe setup provided homogeneous pressure conditions, with negligible differences in the residual resistance: the ~0.1 Ω contribution from the screw–sample interface was nonsignificant, whereas the grain boundaries introduced a small residual resistance in the macroscopic measurements. Data at ambient pressure was obtained after lightly compressing the powder and subsequently reducing the pressure while maintaining a stable current path.

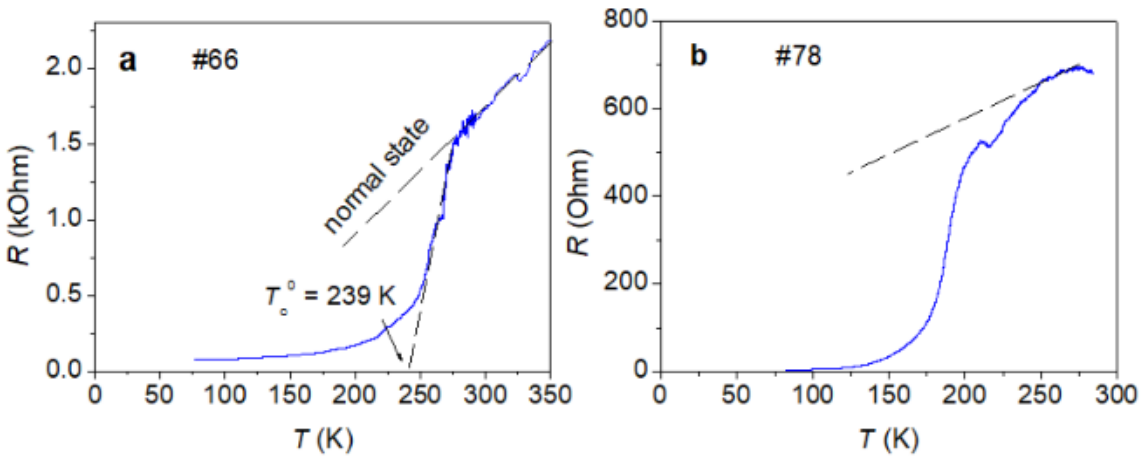

FIG 5. Electrical resistance of lightly compressed CNT@ZSM-5 powder. **a**, Sample without surface treatment with a small residual resistance at low temperatures. The dashed line is an estimation of the normal state resistance. **b**, Sample with surface treatment, where a thin surface carbon layer results in a decrease in the residual resistance.

The electrical resistance results of two batches of samples are shown in Fig. 5. The electrical resistance of the samples substantially decreases. As shown in Fig. 5a, above 280 K, the resistance increases linearly with temperature; however, the resistance sharply decreases below 276 K. At lower temperatures, the resistance remains finite and gradually approaches a residual value of ~1% of the normal-state resistance (above $T_c$) near 100 K. This residual resistance varies across different sample batches, likely reflecting variations in disorder within the CNT structure at the grain surfaces (Suppl. Fig. S7 & S8) [30]. A comparison of the 2-probe technique with a 4-probe technique which does not include any contact resistances (Suppl. Fig. S8e) [30] reveals that the grain boundary resistance causes this difference. The slope of the resistance above the transition increases linearly towards higher temperature typical for a metallic material, thus ruling out a metal to insulator transition as the origin of this pronounced sudden resistance drop. Extrapolation of the steepest slope to zero resistance yields an estimated $T_c^0$ of 239 K, indicating the onset of superconducting phase coherence. This value is consistent with the transitions observed in the magnetic measurements.

To suppress the residual resistance and to investigate a possible superconducting nature of this resistance drop, we changed the sample growth process—by decreasing the annealing duration to 4.5 hours from 5 hours—to intentionally leave a thin carbon surface covering the grains to enhance the electronic coupling between grains. While the transition of this batch occurs at a somewhat lower temperature near 200 K, the low-temperature resistance appears to decrease, in principle, to zero when considered at this large scale (Fig. 5b). Additional data from other batches, including enlarged plots of the raw resistance measurements, are provided in Suppl. Figs. S7 & S8 [30].

### C. Giant pressure effect and superconductivity-like signatures above the ambient temperature

In our setup for the resistance and AC susceptibility measurements, the powder was compacted between two screws, which allowed us to compress the powder with a very small pressure of up to ~0.1 kbar when the screws were tightened with a torque of $M = 1$ Nm using a handheld screwdriver equipped with a torque meter. Such a low pressure is not expected to have a significant effect on typical bulk superconductors; however, we surprisingly observed a very strong effect for all our CNT@ZSM-5 samples. The inset of Fig. 6a shows the same electrical resistance data as shown in Fig. 5a but at an ambient pressure and a finite pressure applied by tightening the screws. The pressure results in a markedly lower overall resistance. The longitudinal force generated by the screw is given by $F = 2\pi \nu M/s$, where $s = 0.25$ is the slope of the M1.2 screw and $\nu = 0.6$ is the efficiency of the screw connection. The resulting pressure, which is on the order of $p \approx 0.1$ kbar, is not strictly hydrostatic because of the absence of a pressure-transmitting medium; however, quasi-hydrostatic conditions are likely produced owing to the fine powder nature of the sample that can easily adjust to the mechanical constraints in the cell. Remarkably, the resistive transition shifts upward by more than 100 K, with the onset of the resistive drop occurring at a temperature greater than the ambient temperature.

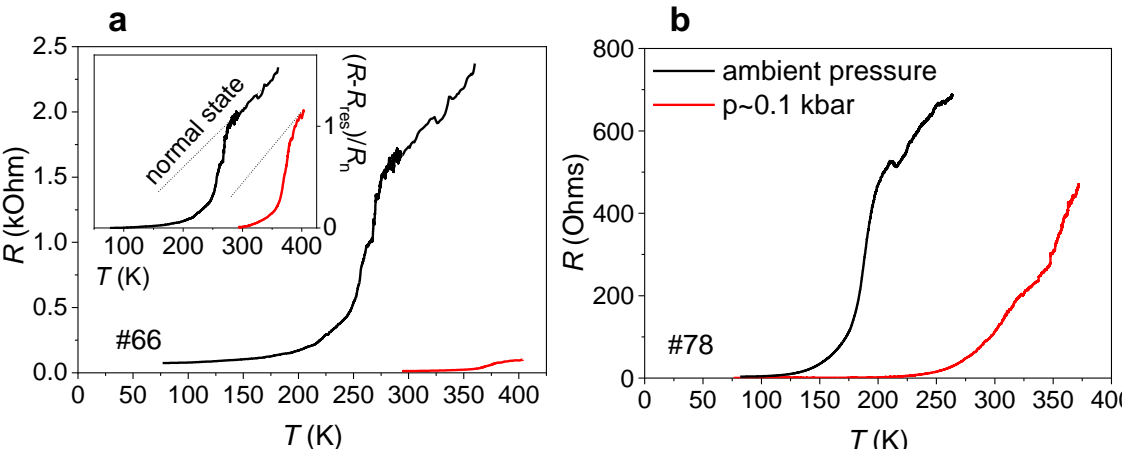

FIG 6. Giant pressure effect of the transition in the electrical resistance. **a**, Early batch of the CNT@ZSM-5 crystallites without surface treatment. The inset show the same data after subtraction of the low temperature residual resistance and normalized by the normal state resistance for clearer comparison. **b**, Most recent surface-treated samples, in which a thin carbon layer was present on the surface, resulting in the vanishing of the low-temperature residual resistance.

To clearly illustrate this transition, we subtracted the small residual resistance at the lowest temperature and normalized the data according to the high-temperature normal state value at the onset of the resistive transition (inset to Fig. 6a). The substantial shift in $T_c$ by approximately 100 K to well above the ambient temperature becomes obvious at this scale, indicating the exceptionally strong pressure dependence of the resistive transition. When the screw was tightened at an ambient temperature prior to the measurement, a reproducible pressure-induced resistance drop was consistently observed across multiple pressure cycles.

We also analyzed a more recent sample batch (#78), for which we attempted to optimize the surface of the CNT@ZSM-5 crystallites by leaving a thin surface carbon layer. For this optimized sample batch, the low-temperature residual resistance vanishes at a pressure of ~0.1 kbar (Fig. 6b), in agreement with the expectations for a superconducting state.

### D. Transition signature in the specific heat

Specific heat as an energy-resolved bulk thermodynamic probe is the ideal tool to investigate the phase transition of superconductors at $T_c$. The specific heat [31] was measured for two batches of samples (#38 and #66) using a few milligrams of the sample, providing similar results confirming reproducibility. The total zero-field specific heat divided by the temperature (*C*/*T*) of CNT@ZSM-5 reveals a second-order phase transition anomaly at 236 K (arrow, Fig. 7a), which is superposed on a large phonon background. The small magnitude of this anomaly compared to the background contribution is anticipated, as the total specific heat is predominantly influenced by phonons at this temperature. The transition temperature coincides with the onset of the Meissner effect and the resistive transition. The specific heat near $T_c$ reveals that the midpoints of the broadened step-like transition differ by approximately 2.5 K across the two samples (sample 1: $T_c$ = 236 K; sample 2: $T_c$ = 233.5 K), with the transition width of sample 2 being significantly broader (inset of Fig. 7a), which we attribute to small variations in doping concentration and homogeneity.

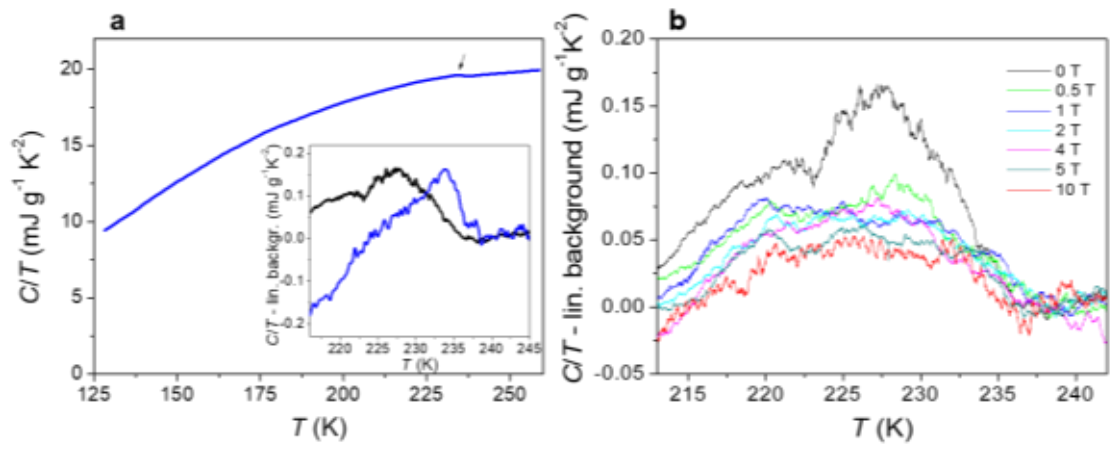

FIG 7. Specific heat of CNT@ZSM-5. **a**, The total specific heat divided by the temperature *C*/*T* under zero field conditions (Sample 1, delineated in blue) shows a 2nd-order phase transition anomaly at 236 K (arrow) attributed to the superconducting transition. The inset shows *C*/*T* for two samples (Sample 1: blue; Sample 2: black) which reveals a slight variation in $T_c$ and the transition width. A linear background fitted above $T_c$ has been subtracted for clarity. **b**, *C*/*T* for Sample 2 under magnetic fields up to 10 T after subtraction of the linear background, showing broadening of the transition due to fluctuations.

Accurately estimating the phonon contribution to the specific heat is a complex task for high-temperature superconductors, with few approaches available for estimating the electronic specific heat in cuprate superconductors. This task is even more challenging for the CNT-zeolite composite owing to its complex nature. This prevents us from the extraction of superconducting parameters normally accessible in classical superconductors. Similar to the case of cuprates, we therefore are forced to focus on the interpretation of the shape of the transition anomaly and its field dependence. Fig. 7b presents *C*/*T* data of sample 2 measured under various applied fields up to 10 T, following the subtraction of a linear background solely to enhance clarity of the presentation. The zero-field data exhibit a small but relatively sharp jump centered at 233.5 K, followed by a broader bump peaking at 228 K. We cannot determine whether this represents the typical BCS mean-field type phase transition, which is characterized by a jump in the specific heat, or the diverging lambda-shaped transition typical of cuprate superconductors, where 3D–XY phase fluctuations dominate the transition shape [32,33], owing to the width of this feature. The zero-field transition of sample 1 (Fig. 7b) shows a small increase below the jump that may indicate a lambda transition.

Under a magnetic field, the anomaly of sample 2 broadens rapidly, resulting in a broad bump. This behavior is similar to the fluctuation-dominated superconducting transition observed in cuprates, as the field reduces the effective dimensionality of a superconductor, driving the transition toward a continuous crossover [34] in contrast to the superconducting phase transition of classical superconductors that happens at a sharply defined temperature. Phase coherence is then established only below a vortex melting transition that occurs at lower temperatures [35]. Given the unknown value of the electronic specific heat, which cannot be determined at such high temperatures without knowing the phonon contribution, more detailed information, such as the coupling strength, paring symmetry, energy gap value and condensation energy, cannot be extracted, but the observed phase transition anomaly—reminiscent of that in cuprate superconductors—further supports the formation of a superconductivity-like electronic condensate.

### E. Energy gap and particle hole symmetry in point–contact spectroscopy

Point–contact spectroscopy is an energy-resolved technique that has been proven useful to provide information on superconducting pairing gaps and order parameter symmetry. We thus use it to gain further evidence of a superconductivity-like electronic condensate. With our scanning probe, we can vary the tunneling barrier height or transparency by adjusting the pressure of a sharp tip pressed against the sample, allowing us to tune the contact between the tip and a superconductor between the high-transparency Andreev limit and the low-transparency tunneling limit. For a superconductor, the resulting spectra can be analyzed by the Blonder–Tinkham–Klapwijk (BTK) formalism [36], which allows us to extract fundamental parameters - most importantly - the superconducting energy gap, which we apply here to test a possible superconducting origin of observed features. In the tunneling limit, the spectra exhibit characteristic gap suppression with two coherence peaks or shoulders, whereas in the Andreev limit, a plateau appears within the bias voltage range corresponding to the gap size. Such apparent features superposed on a smooth parabolic background (Fig 8a & Suppl. Fig. S9) [30] were observed below ~240 K.

The signal-to-background ratio varied significantly across the contacts we established. The background was predominantly metallic, as the few-micron-sized insulating zeolite crystallites hosting the metallic CNTs prevented standard scanning tunneling experiments. To address this, we coated the crystallites with a thin Au layer, aiming to hold them in place while probing a proximity-induced superconducting gap. However, due to the mechanical instability of the crystallites, we had to pierce the tip through the Au layer directly into the zeolite to establish a point contact with the CNTs, while the Au layer still contributed a metallic background to the electrical conductance. In Fig. 8a, we present data showing a gap-like depression at low temperature (black data), accounting for 25% of the total signal, a reasonably large effect. Unfortunately, the thin Au film on this sample was not mechanically stable enough to hold the crystallites in place at higher temperatures. Thus, we studied the temperature dependence on a sample with a thicker Au layer, which reduced the signal-to-background ratio strongly (see Suppl. Fig. S9 for the raw data [30]) but still provided sufficient resolution to observe very similar features resembling superconducting tunneling characteristics.

We removed the parabolic background by normalizing the data via a fitted parabola. Note that this does not affect the overall shape and interpretation of the data but only serves for clarity of the presentation and to improve the BTK fits. Examples of normalized data are shown in Fig. 8b–e for 4 selected temperatures. The curves were obtained at different temperatures and show new contact points on the sample, which have distinct contact transparencies. This does not affect the analysis, as the contact transparency is a parameter in the BTK model that can be adjusted to capture the nature of the contact, while the energy gap size Δ is independent of it and the parameter of interest. The data taken over the entire temperature range up to the ambient temperature reveal that within the resolution limit, the spectra are symmetric with approximately zero bias (Fig. 9a), reflecting the particle–hole symmetry characteristic uniquely known for superconductors.

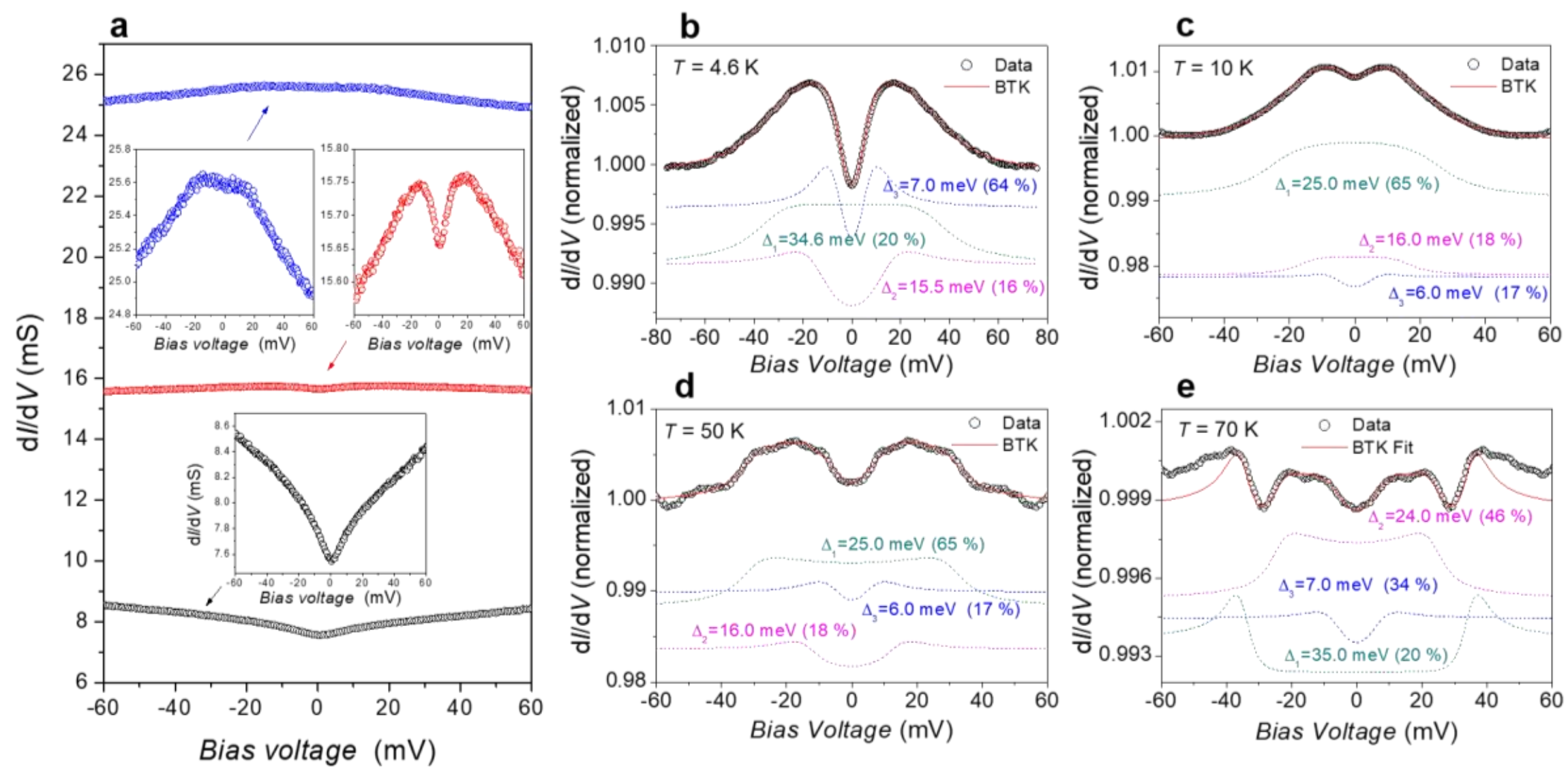

FIG 8. **a** Examples of raw data of 3 point–contact spectra acquired below 10 K, with contact transparencies varying from the low-transparency tunneling limit (black) over an intermediate range (red) to the high-transparency Andreev limit (blue). Due to a very thin Au coverage in this device, the signal to background ratio was up to 25 % of the total signal (black curve). The insets show enlargements of the 3 spectra. The difference was achieved by varying the tip distance from the sample surface. The ability to tune the spectra between these two limits supports a possible superconducting origin of the spectra. **b-e**: Signature of superconducting-like gaps in selected normalized point–contact spectra of a device with thicker Au layer at 4.6 (**a**), 10 (**b**), 50 (**c**) and 70 K (**d**) (circles) with respective BTK fits (red solid lines). The data have been symmetrized to increase the relative resolution. The additional dashed lines indicate the individual contributions in our 3-gap model, which are attributed to 3 bands crossing the Fermi level originating from the (2,1) and (3, 0) CNTs.

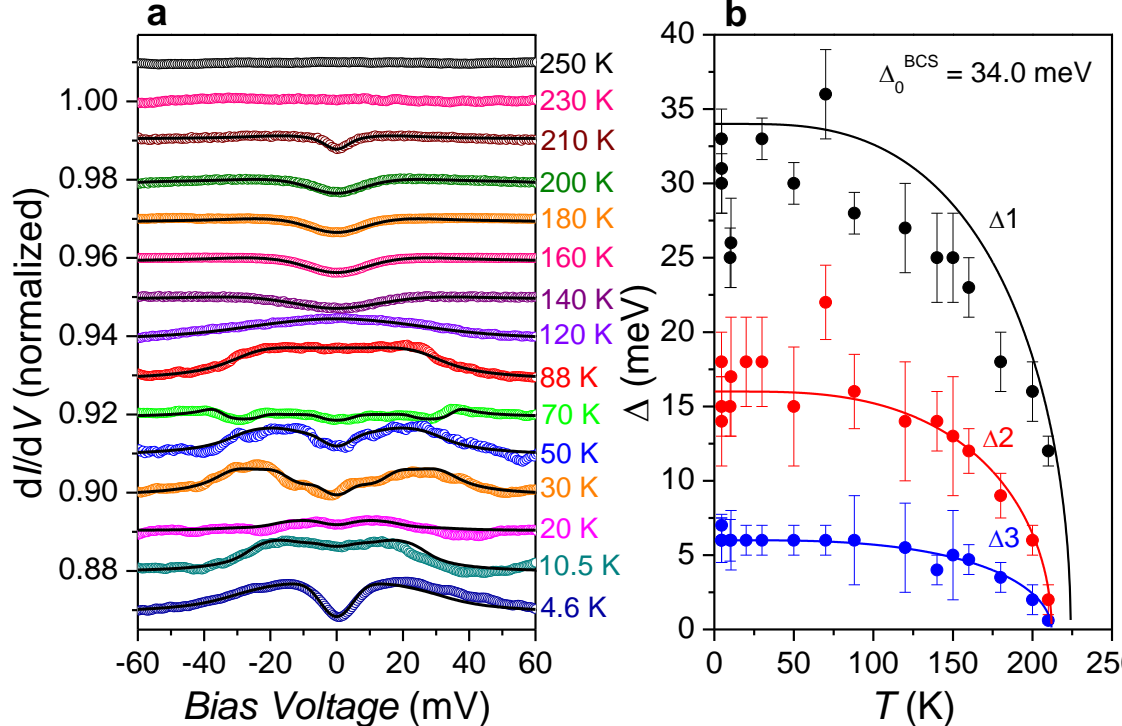

FIG 9. Temperature dependence of energy gaps in the point–contact spectra of CNT@ZSM-5. **a**, Temperature dependence of the normalized original (non-symmetrized) point–contact spectra up to 250 K. Offsets of 0.01 have been added, except for the 230 K data. The gap features vanish above 230 K. The BTK fits are included as solid lines. **b**, Temperature dependence of three superconducting gaps from the BTK fits. The error bars are derived from the uncertainty of the BTK fits. The largest gap closely follows the BCS prediction of $T_c$ = 224 K ($\Delta_0^{BCS} \approx$ 34.0 meV), and the smaller gaps show similar trends with lower $\Delta_0$ values. The data scatter reflects variations in the contact conditions.

The curves cannot be fitted by a standard single-gap s-wave model owing to the additional structures in the spectra, which are particularly evident for the 70 K data. The spectrum at this temperature shows a large gap-like depression, and a plateau-like structure with another small depression near zero bias can be observed within this large depression. Good agreement is obtained using a 3-gap s-wave model, where multiple electronic bands are involved in the condensate, which accounts perfectly for all data taken at different temperatures.

In the extreme tunneling limit, one notable feature is the absence of the so-called coherence peaks, which typically frame a superconducting gap. However, the BTK model can simulate spectra without coherence peaks if the broadening factor associated with the quasiparticle lifetime is large [36]. In our data, this absence may have occurred owing to disorder at the sample surface (note that owing to the powder nature of the sample, the samples were measured as grown, without any cleavage or surface cleaning) or a quasi-one-dimensional nature of the condensate.

At temperatures above 120 K, the Andreev reflection contribution from the parabolic background became increasingly difficult to distinguish when measuring high-transparency contacts; therefore, we chose to tune the contact to the low-transparency tunneling limit, where the gap depression was clearly visible. At lower temperatures, the high-transparency Andreev limit was preferable because the distinct features of the 3 gaps were more evident. At temperatures of 230 K and above, the data became perfectly parabolic, irrespective of the contact transparency, indicating the absence of coherent superconductivity.

We next plotted the temperature dependence of the fitted gaps to compare it to the expected behavior proposed by the standard BCS model of superconductivity. The largest leading gap varies at low temperatures between 25 and 33 meV; decreases at temperatures above ~70 K; and appears to vanish at temperatures below approximately 230 K (Fig. 9b). This large variation likely occurred because we needed to guide the tip to the sample for each measurement; thus, slightly different sample positions may have been probed. The exact tip pressure may also play a certain role, considering the large pressure dependence of the transition temperature observed in the resistance measurements. In the figure, the black line represents the estimation for a 224 K superconductor on the basis of BCS theory, which corresponds to a gap value $\Delta_0$ of 34 meV at $T$ = 0 K:

$$\Delta(T) \approx 1.76 k_B T_c \tanh\left(1.74\sqrt{\left(\frac{T_c}{T}-1\right)}\right) = \Delta_0 \tanh\left(1.74\sqrt{\left(\frac{T_c}{T}-1\right)}\right) \quad (1)$$

This value is slightly larger than the average of our determined values below 20 K but is reasonably close to our estimate. The low-temperature limit of the second gap is 14–17 meV, and the low-temperature limit of the smallest gap is 6 meV. The two smaller gaps appear to close at a slightly lower temperature of 212 K but are reasonably close to the 224 K temperature determined for the largest gap within the resolution limit. This behavior is similar to that observed in known multigap superconductors [37].

In Suppl. Fig. S10 [30], we show point–contact data measured at 4.6 K under zero-field conditions and at 15 T to investigate the magnetic field dependence of the energy gaps. The contact needed to be readjusted following the field sweep to 15 T, resulting in the 15 T data being within the Andreev regime. However, within the resolution limit, the edges of the gap do not vary substantially, suggesting an extremely high upper critical field far beyond 100 T, a characteristic that is extremely attractive for technological applications in the form of high magnetic field solenoids.

## V. DISCUSSION

We observed the features that are commonly expected for superconductors across five complementary measurements that are commonly used to study superconductivity, including thermodynamic measurements such as magnetization and specific heat, electrical transport measurements and energy-resolved point–contact spectroscopy measurements. All the observed features are clearly consistent with those of superconductors and together allow little room for any interpretation that does not involve an electronic condensate; however, the results are observed at significantly higher temperatures than any confirmed ambient-pressure superconductor. In the following we discuss it in the framework of a potential new high temperature superconducting phase, or superconducting-like phase.

The DC magnetization response was very small, resulting in incomplete Meissner screening. Additionally, while the lower critical field ($H_{c1}$) is expected to be small, a larger applied field is needed to observe minor magnetization contributions, which further reduces the Meissner screening effect. The difference between the ZFC and FC branches is evident but varies depending on the sample batch and field strength. Some data show diamagnetic contributions in both branches, whereas in rare cases, the FC curve even shows a positive contribution. This phenomenon is known as the paramagnetic Meissner effect and usually is attributed to a superconducting surface effect [38]. All the ZFC curves demonstrate the Meissner effect, albeit not in the form of a sharp jump but rather as a continuous kink. This behavior can be attributed to the small signal magnitude, which prevents detection of diamagnetism in the case of very small fields below $H_{c1}$. The applied field needed to be at least 100 Oe to clearly observe a Meissner-like response, which typically results in a continuous transition.

To understand the smallness of the observed DC magnetization features, one needs to consider the filamentary network structure comprising thin CNTs with radius $a$, the dimensionless magnetic susceptibility scales as $-C\left(\frac{a}{\lambda}\right)^2$, where $\lambda$ denotes the magnetic penetration length and $C$ is a pre-factor that is $O(1)$ [28]. For $a \approx 1.25\ A$ and $\lambda \approx 50\ nm$, we have , $\left(\frac{a}{\lambda}\right)^2 = 6.25 \times 10^{-6}$. Such a small susceptibility is consistent with an extremely large effective $\lambda_{eff}$ for the filamentary network system. When the sample thickness $d$ is comparable to or smaller than $\lambda_{eff}$, the system reaches the Pearl limit [39], under which the susceptibility is $\chi = -\left(1 + \frac{2\lambda_{eff}}{d}\right)^{-1}$. Consequently, a large discrepancy between the SQUID and AC susceptibility Meissner measurements may arise from the geometric difference between the thin loose powder layer used in the SQUID measurements and the compressed solid cylinder used in the AC susceptibility measurements. That latter can exceed 1 mm; while the compact sample is also expected to have a smaller $\lambda_{eff}$ than that of the loose powder.

We have demonstrated that upon optimization of the grain boundaries of compressed powder samples, a zero-resistance state can be approached below at least 70 K at ambient pressure, which can be enhanced to at least 200 K upon application of weak pressure. The main resistive jump which marks the transition occurs for almost all batches near the temperature below which a Meissner effect and an energy gap is observed. It can be shifted to well above ambient temperature by light pressure. The latter could mark a formation of Cooper pairs, which occurs at temperature above the zero-resistance state due to the filamentary quasi-1D nature of the material. The zero-resistance temperature can likely be further increased to ambient temperature upon application of higher pressure which drives the material to a more 3D condensate.

The relatively large size of the specific heat anomaly at $T_c$ suggests that a significant volume fraction of the samples exhibit an electronic condensate. Similar to the behavior observed in cuprates, the transition appears to be fluctuation dominated, and the presence of a magnetic field broadens the transition into a bump-like crossover. In these materials, this effect arises because of the magnetic field-induced finite size effect in the presence of fluctuations of the anisotropic 3D-XY universality class [40,41]. The associated length scale, which is determined by the vortex separation, prevents divergence of the correlation length when approaching $T_c$. Unlike cuprates, our sample consists of a 3D network of 1D metallic objects, whereas cuprates consist of weakly coupled 2D layers. The reduced dimensionality strongly increases fluctuations in the phase of the order parameter of the condensate. In this case, Cooper pairs would first form at a high temperature in the 1D nanotubes (or in the 2D planes of cuprates), with no phase coherence observed. Afterward, correlations gradually form until, at $T_c$, a phase-coherent condensate is formed. For cuprates, the fluctuations can extend up to twice the $T_c$ value and can be observed in the thermodynamic quantities [32–34] or as vortex excitations via the Nernst effect [42]. A similar situation is likely observed here, as the magnetization and specific heat results suggest that our material could be a high-$T_c$ material with even higher $T_c$ of 220–250 K. The width of this transition would correspond then to the range in which Cooper pairs begin to develop correlations until a phase-coherent zero-resistance state is formed. Notably, Cooper pairs may exist at an even higher temperature in the form of preformed pairs [43,44], but these pairs lack correlations.

Among the measurements performed, point–contact spectroscopy provided evidence of an electronic condensate up to about 230 K. The spectra demonstrate particle–hole symmetry, a characteristic feature of superconductors. The Andreev limit further suggests a superconducting origin of the measured effects. Three energy gaps were identified. To understand the origin of these gaps, we performed ab initio band structure calculations. The electronic band structures of the (2,1) and (3,0) CNTs are shown in Suppl. Fig. S1 [30], confirming their metallic nature. The presence of two types of CNTs is revealed by the Raman signals and ab initio calculations, as discussed in the Methods section. This finding is also consistent with the geometric constraint provided by the two sets of pore channels in ZSM-5. The ab initio calculations also reveal that boron doping downshifts the Fermi level by ~0.5 eV, placing this level in the direct vicinity of a van Hove singularity for the (2,1) CNTs (Suppl. Fig. S1) [30]. We thus believe that the (2,1) CNTs are the leading cause of the observed superconductivity-like features. The Fermi level crosses three bands, which are likely responsible for the three energy gaps observed. The (3,0) CNTs may also contribute via the proximity effect. Despite the scatter in the measured gap values, which resulted from the necessary adjustments to the tip to perform each temperature measurement, the obtained data are consistent with BCS theory for a superconductor with a $T_c$ near 224 K. However, within the limits of the error bars, the temperature dependence of the gaps appears to decrease more continuously and linearly than strictly following the BCS trend, possibly due to a stronger electron-phonon coupling than considered in the BCS model.

One of the most striking observations is the strong effect of pressure on $T_c$. The substantial increase in $T_c$, as observed in the resistance measurements, to values exceeding room temperature is highly unusual: $T_c$ is increased by ~100 K under a pressure of ~0.1 kbar, whereas typically several orders of magnitude higher pressures are required to change $T_c$ values of superconductors by only 1 K. Although our experiment was not planned as a high-pressure experiment with non-ideal conditions, the significant increase in $T_c$, with the superconductivity-like features induced at an ambient temperature by simply turning a small screw with a handheld screwdriver, is remarkable and could have wide-ranging technological applications, either for highly sensitive pressure sensors, or simply through the fact of having a highly efficient control parameter to widely tune the superconducting characteristics for new types of electronic applications. Further experiments under better hydrostatic conditions with a controlled pressure medium and with a precise pressure gauge are desirable to investigate this pressure dependence more quantitatively.

The strong pressure dependence may stem from the vicinity of the Fermi level near the van Hove singularity, as revealed by our bandstructure calculations (Suppl. Fig. S1) [30]; pressure is likely to shift the associated flat band, enhancing the density of states at the Fermi level—a key factor for achieving a high $T_c$ in a superconductor according to the McMillan formula [3]. Additionally, the soft zeolite matrix may deform the CNTs under pressure, altering the phonon modes and influencing $T_c$ through a phonon-mediated mechanism. Note that the (2, 1) CNTs are extremely unstable and are stabilized only by their confinement within the zeolite matrix, as indicated by the imaginary phonon frequencies in the phonon density of states (Suppl. Fig. S1) [30]. This result, together with the vicinity of the van Hove singularity, makes these (2, 1) CNTs extremely exotic metallic materials with parameters rarely found elsewhere. However, we cannot determine the pairing origin from the current data, which stimulates future experiments, such as muon spin resonance experiments.

Our interconnected network of CNTs has unique features that could further explain the strong pressure effect. The perpendicular CNTs are separated by a small gap of only 1.3 Å (Fig. 1e) and do not intersect. At such small separations, some bonding may be inevitable. However, such intersections would be mechanically weak points in the system. Owing to the soft crystal lattice of the ZSM-5 zeolite material, pressure deforms the zeolite, thereby pushing the perpendicular CNTs closer until they eventually touch. This leads to a dimensional crossover between 1D and 3D from individually behaving 1D elements with a filamentary condensate towards a 3D coupled phase-coherent bulk state. As previously discussed, within a superconducting interpretation, phase-incoherent, preformed Cooper pairs may exist well above $T_c$, the temperature at which they condense into a phase-coherent condensate. When the CNTs approach their perpendicular intersections, the resulting 3D network will suppress fluctuations, resulting in the establishment of a phase-coherent condensate at a higher temperature. However, further experiments are needed to confirm the existence of phase-incoherent pairing above $T_c$, including more sensitive tunneling experiments and measurements of the thermoelectric Nernst effect, with the latter having proven highly sensitive in the case of cuprates [42].

## VI. CONCLUDING REMARKS

We have investigated boron doped ultrathin carbon nanotubes confined within the ~5 Å channels of ZSM 5 zeolite, forming a three-dimensional network of mechanically coupled one-dimensional conductors. Across five independent experimental probes—electrical transport, DC and AC magnetic susceptibility, specific heat, and point contact spectroscopy—we observe mutually consistent signatures of a phase coherent electronic state emerging between 220 and 250 K at ambient pressure. These signatures include a sharp resistive drop, diamagnetic screening, a reproducible thermodynamic anomaly, and particle–hole symmetric gap like spectra exhibiting Andreev reflection. While each measurement individually admits alternative interpretations, their concurrence at a common temperature scale across multiple batches provides strong evidence for collective electronic behavior with superconducting-like characteristics.

An additional and striking observation is the pronounced sensitivity of this state to modest applied pressure. Pressures below 0.1 kbar shift the resistive onset upward by nearly 100 K and modulate the room temperature resistance by more than three orders of magnitude. This giant pressure response is consistent with the fragile mechanical coupling between ultrathin CNTs and suggests that the system lies near a 1D–3D crossover in intertube coherence. Regardless of its microscopic origin, this tunability represents a technologically relevant property of the CNT@ZSM 5 composite.

Taken together, these findings identify boron doped ultrathin CNT networks as a promising platform for exploring high temperature phase coherent states in low dimensional carbon systems. Further work is required to fully elucidate the pairing mechanism, quantify the role of confinement and doping, and establish the extent to which the observed behavior reflects true superconductivity. Nonetheless, the present results demonstrate that confinement engineered CNT networks can host collective electronic phenomena at unexpectedly high temperatures and motivate deeper investigation of their structural, electronic, and mechanical properties.

## VII. METHODS

### A. DC magnetization

DC magnetization was measured with a commercial Quantum Design® Vibrating Sample SQUID magnetometer which allows high relative resolution measurements of $10^{-9}$ – $10^{-8}$ emu. Temperature dependent data was taken under zero-field cooled (ZFC) and field cooled (FC) conditions upon stabilizing the temperature with averaging times of 2 – 10 seconds. The sample was attached to the sample holder using highly diluted Model VGE7031® insulating varnish / adhesive (LakeShore Cryotronics®). The glue has slightly paramagnetic properties, which can be minimized through strong dilution with ethanol. We also attempted using different types of vacuum grease, which showed a smaller almost temperature independent background contribution, but the transition to a solid phase caused some unwanted anomalies. To manipulate the raw data as little as possible, we used linear background fits in a relatively small temperature interval to remove this background. Raw data is shown in Suppl. Figs. S2 – S4 [30]. The background stems partially from a paramagnetic background from the zeolite and glue used to mount the sample, as well as from an unwanted small linear signal drift of the SQUID magnetometer.

### B. Specific heat

The specific heat was measured with a dedicated home-made alternating temperature (AC) calorimeter with unique design where a sapphire sample chip with Joule heater deposited to its back is suspended on a thermopile made from 8 AuFe0.07% / Chromel thermocouples [31]. The sample temperature was modulated at a frequency of ~1.4 Hz with a temperature modulation kept below 1 mK amplitude of 1 mK. The thermocouple voltage was amplified by $10^4$ times via a Model A20 DC Nanovolt amplifier (EM Electronics®), which is not impacted by $1/f$ noise, and input into a Model SR830 digital lock-in amplifier (Stanford Research Systems®). The chip was attached to a copper block with an additional thermometer, which served as a thermal bath. The calorimeter was enclosed in a sealed vacuum chamber with a thermal link to the environment and inserted in a $^4$He variable temperature inset of a superconducting magnet cryostat. The addenda of the heat capacity of the chip was determined separately and removed. This unique design allows very high relative resolution measurements with $\Delta C/C$ values as low as $10^{-4}$. Data were collected at steps of 1 mK at a cooling rate of 0.1–0.5 K/min between ambient temperature and ~100 K with a high data point density.

### C. Electrical resistance and AC susceptibility

The electrical resistance was measured using macroscopic samples consisting of approximately 1 mg of powder. Because the powder lacked sufficient cohesion, we fabricated an insulating polyamide body with a 1.2 mm screw hole drilled completely through its length (Fig. 10a). The powder was filled into the hole and compressed between two screws, forming a conducting sample in which the screws served as terminals for two-probe measurements. We also adopted an improved design featuring two screw holes intersecting at a 45-degree angle, allowing the sample to be compressed by four screws, which served as terminals for four-probe experiments (Fig. 10b). Because the superconductivity in CNT@ZSM-5 is extremely sensitive to the pressure applied by the screws, the two-probe design allowed a simple pressure study of the superconducting transition, where the pressure was roughly calculated on the basis of the measured torque applied to the screws. For the four-probe device, we attempted the same experiment (Suppl. Fig. S8); however, the design caused significant pressure gradients, resulting in a large broadening of the superconducting transition. We used a Model 6221 AC and DC (Keithley®) current source in combination with either a digital Stanford Research Systems SR850® lock-in amplifier for AC measurements or an Agilent 34401A® digital multimeter with a current reversal technique to avoid thermoelectrical voltages for DC measurements. Both methods resulted in equally good results.

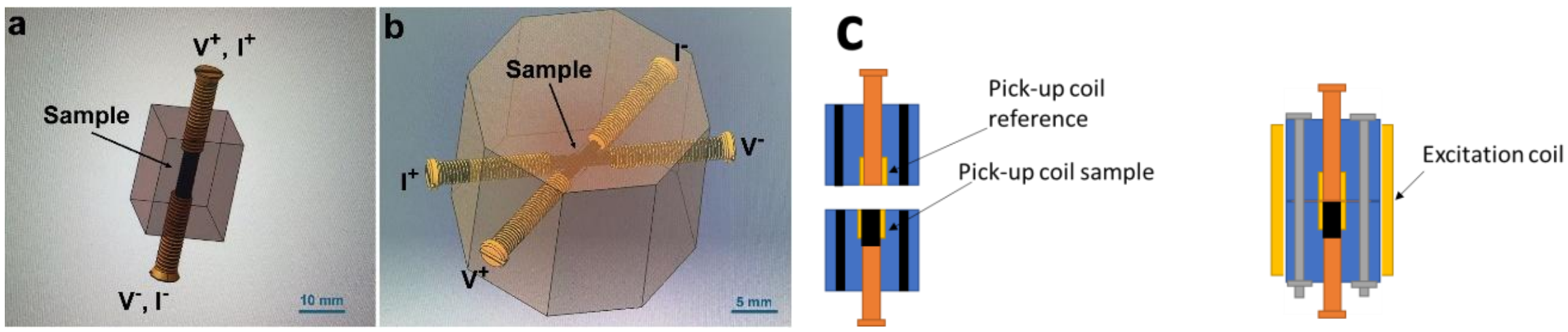

FIG 10. Screw devices for measuring the electrical resistance of the powder sample. The sample is placed in the center of an M1.2 screw hole that traverses through the entire length of a polyamide body; the sample is compressed between one pair of screws **a**, or two pairs of screws in screw holes intersecting at a 45-degree angle **b**. The pairs of screws serve as electrical terminals for either two-probe measurements **a** or four-probe measurements **b**. The two-terminal device **a** could also be used as a simple pressure cell capable of applying pressures of up to ~0.1 kbar. **c**, Modified 2-probe device for simultaneous measurements of the AC susceptibility. The device in **a** was separated into two pieces (left), and a pair of small compensated pickup coils were integrated into the screw holes. After reassembly, an external excitation coil was added.

Initially, we used a $^4$He cryostat with a variable temperature insert; however, we found that the required temperature range extended beyond ambient temperature, which was inaccessible with the low-temperature device. Therefore, we switched to a simpler experimental setup in which the device was mounted in a closed, solid copper pot containing an integrated, calibrated Pt1000 thermometer. The copper pot served as a thermal bath. The sample stage was wrapped in adhesive copper tape to establish good thermal contact with the thermometer and the thermal bath. The measurement protocol consisted of first cooling the pot with liquid nitrogen to 77 K and then performing measurements while the pot was slowly warmed to ambient temperature. These results were in perfect agreement with those obtained using the cryostat. We subsequently used a hot plate to heat the pot to 400 K at a rate of ~1 K/min.

To enable simultaneous measurements of the electrical resistivity and AC susceptibility, we modified the original two-probe device. The device was cut into two parts, and two 4-turn, 1-mm-diameter pickup coils were inserted into the existing screw holes (Fig. 10c). The two halves were then reassembled and secured with additional screws. One pickup coil was loaded with the sample, while a brass screw was inserted into the second coil to act as a reference. The two parts were subsequently pressed firmly together and fixed in place.

As in the original configuration, the brass screws entering from both sides served as electrodes for the two-probe resistivity measurements, and pressure was applied to compact the sample. After reassembly, an external excitation coil with approximately 500 turns was wound around the sample.
For the AC susceptibility measurements, a digital Stanford Research Systems SR850® lock-in amplifier was used. The best results were obtained at relatively high working frequencies near 1 kHz.

### D. Point–contact spectroscopy

The point–contact spectroscopy data were obtained using a custom scanning tunneling probe. Many CNT@ZSM-5 crystallites were densely packed on a conducting cupper substrate and subsequently covered through evaporation by a very thin (~5 Å) gold film, which served to supply electrical contact near the tip rather than passing the current through the crystallites from the back of the substrate. The objective was to search for a proximity-induced superconducting gap in the gold layer via tunnelling spectroscopy; however, better results were obtained when the tip was pierced through the gold layer to directly contact the CNT@ZSM-5 crystals in point–contact mode. Data were collected at different contact resistances between 10 Ohms and a few kOhms between the tip and the CNT@ZSM-5 crystals; thus, the barrier height was varied so that superconductivity can be investigated under both the low-transparency tunneling limit and the high-transparency Andreev limit [36,45], which provide distinct responses.

To perform the point–contact tunneling and Andreev reflection spectroscopy measurements, we employed a Keithley 6221® source capable of delivering both AC and DC currents in conjunction with an Agilent 34411A® digital multimeter and an SR830 lock-in amplifier. The measurement protocol involved varying the DC incrementally in small steps from negative to positive values. This DC sweep established the bias voltage across the junction. Simultaneously, a small-amplitude AC that was comparable in magnitude to the DC step size was superimposed to enable lock-in detection of the differential conductance, d*I*/d*V*, as a function of the applied bias. A digital multimeter was used to monitor the resulting voltage. This current-driven approach is particularly well suited for probing point-contacts. The measurements were done in a $^3$He insert within a 15 T superconducting magnet cryostat which was optimized for vibrational damping required in such experiments.

### E. Supporting ab-initio calculations

We employed first-principle density functional theory method to investigate the electronic band structure and the phonon spectrum of (2,1) and (3,0) CNTs. A mixed-basis pseudo-potential code [46,47] is used to perform the calculations. We use plane wave with 24 Ryd cutoff energy and 2s- and 2p-like localized functions to describe the valance electrons of carbon, and the Brillouin zone sampling in the band structure and phonon spectrum calculations are set to 1×1×72 k-points for (2,1) CNT and 1×1×256 k-points for (3,0) CNT, respectively. The atomic structures are fully relaxed before the band structure and phonon spectrum calculations. The results can be found in Suppl. Fig. S1. The results show the (2,1) CNT and (3,0) CNT to be metallic, and boron doping moves down the Fermi level of (2,1) CNT to the vicinity of the van-Hove singularity. This is favorable to introduce electron-phonon interaction related phenomena, like phonon mediated superconductivity and Peierls distortion. As the CNTs are embedded in a three-dimensional network, Peierls distortion is suppressed. We deduce that boron doping is likely to introduce superconductivity in our case. Phonon spectra show that individual (3,0) is robust against Peierls distortion, while individual (2,1) CNTs are unstable. Phonon frequency softening due to electron–phonon coupling can clearly be observed in the phonon spectrum of the (3,0) CNT.

Note that the confinement of the zeolite framework is not considered in this ab initio calculation because such a system is beyond our computational capacity. However, the confinement of the zeolite framework is very important for stabilizing the (2,1) CNTs and therefore plays a crucial role in the emergence of superconductivity. Additionally, the phonon spectra of both types of CNTs are calculated for the undoped case. Electron–phonon coupling scatters electrons around the Fermi level; thus, the variation in the Fermi level is crucial to the phonon spectrum and, consequently, the phonon-mediated electron–electron interactions. These questions remain to be investigated in the future.

**ACKNOWLEDGMENTS**
We thank U. Lampe and K. Paillot for their technical support. PS would like to acknowledge the support from the HKUST Jockey Club Institute for Advanced Study and thank the Shun Hing Education and Charity Fund, especially the late Dr. William Mong, for their sustained support over two decades. RL thanks KT Law for their fruitful discussions on the particle–hole symmetry of the point contact spectra and CT Chan for his financial support. RL acknowledges SK Goh, M. Leroux and C. Simon for their fruitful discussions.

**DATA AVAILABILITY**
The data that support the findings of this study are available at [to be added].

**Supplemental Materials (ref. [30])**
The Supplementary Materials contain additional data, including band structure and phonon modes from ab-initio calculations (Suppl. Fig. S1), DC magnetization raw data without background subtraction, DC magnetization of additional sample batches (Suppl. Figs. S2-S6), additional resistance data at various pressures and from different batches under 2-probe and 4-probe configuration for comparison (Suppl. Figs. S7, S8), as well as additional point contact spectroscopy data, including raw data without background subtraction and magnetic field dependence (Suppl. Fig. S9-S10).

## REFERENCES

1. Hepting M., Room-temperature superconductivity: extraordinary claims require extraordinary evidence, Nature 62 (2023) 475–476. https://doi.org/10.1038/d41586-023-02857-2.
2. Onnes H.K., The Resistance of Pure Mercury at Helium Temperatures, Springer, Dordrecht, 1911.
3. McMillan W.L., Transition temperature of strong-coupled superconductors, Phys. Rev. 167 (1968) 331–344. https://doi.org/10.1103/PhysRev.167.331.
4. Bardeen J., Cooper L.N., Schrieffer J.R., Theory of superconductivity, Phys. Rev. 108 (1957) 1175–1204. https://doi.org/10.1103/PhysRev.108.1175.
5. Bednorz J.G., Müller K.A., Possible high Tc superconductivity in the Ba−La−Cu−O system, Z. Phys. B Condens. Matter 64 (1986) 189–193. https://doi.org/10.1007/BF01303701.
6. Superconductivity at 93 K in a new mixed-phase Y-Ba-Cu-O compound system at ambient pressure, Phys. Rev. Lett. 58 (1987) 908–910. https://doi.org/10.1103/PhysRevLett.58.908.
7. Schilling A., Cantoni M., Guo J.D., Ott H.R., Superconductivity above 130 K in the Hg–Ba–Ca–Cu–O system, Nature 363 (1993) 56–58. https://doi.org/10.1038/363056a0.
8. Stewart G.R., Superconductivity in iron compounds, Rev. Mod. Phys. 83 (2011) 1589–1652. https://doi.org/10.1103/RevModPhys.83.1589.
9. Wang B.Y., Lee K., Goodge B.H., Nickelate superconductors: an emerging platform for superconductivity research, Annu. Rev. Condens. Matter Phys. 15 (2024) 305–324. https://doi.org/10.1146/annurev-conmatphys-032922-093307.
10. Sun H., Huo M., Hu X., Li J., Liu Z., Han Y., Tang L., Mao Z., Yang P., Wang B., Cheng J., Yao D.-X., Zhang G.-M., Wang M., Signatures of superconductivity near 80 K in a nickelate under high pressure, Nature 621 (2023) 493–498. https://doi.org/10.1038/s41586-023-06408-7.
11. Ashcroft N.W., Metallic hydrogen: A high-temperature superconductor?, Phys. Rev. Lett. 21 (1968) 1748–1749. https://doi.org/10.1103/PhysRevLett.21.1748.
12. Zhang S., Zhang M., Liu H., High-temperature superconductivity in hydrogen-rich materials: a review, Appl. Phys. A 127 (2021) 684. https://doi.org/10.1007/s00339-021-04802-4.
13. Drozdov A.P., Eremets M.I., Troyan I.A., Ksenofontov V., Shylin S.I., Conventional superconductivity at 203 Kelvin at high pressures in the sulfur hydride system, Nature 525 (2015) 73–76. https://doi.org/10.1038/nature14964.
14. Drozdov A.P., Kong P.P., Minkov V.S., Besedin S.P., Kuzovnikov M.A., Mozaffari S., Balicas L., Balakirev F.F., Graf D.E., Prakapenka V.B., Greenberg E., Knyazev D.A., Tkacz M., Eremets M.I., Superconductivity at 250 K in lanthanum hydride under high pressures, Nature 569 (2019) 528–531. https://doi.org/10.1038/s41586-019-1201-8.
15. Iijima S., Helical microtubules of graphitic carbon, Nature 354 (1991) 56–58.
16. Oberlin A., Endo M., Koyama T., Filamentous growth of carbon through benzene decomposition, J. Cryst. Growth 32 (1976) 335–349.
17. Dresselhaus M.S., Dresselhaus G., Saito R., Physical Properties of Carbon Nanotubes, World Scientific, Singapore, 1998.
18. Haruyama J. (Ed.), Carbon-based Superconductors: Towards High-Tc Superconductivity, Pan Stanford Publishing, Singapore, 2014. https://doi.org/10.1201/b15672.
19. Hebard A.F., Rosseinsky M.J., Haddon R.C., Murphy D.W., Glarum S.H., Palstra T.T.M., Ramirez A.P., Kortan A.R., Superconductivity at 18 K in potassium-doped C60, Nature 350 (1991) 600–601. https://doi.org/10.1038/350600a0.
20. Shi W., Wang Z., Zhang Q., Zheng Y., Ieong C., He M., Lortz R., Cai Y., Wang N., Zhang T., Zhang H., Tang Z., Sheng P., Muramatsu H., Kim Y.A., Endo M., Araujo P.T., Dresselhaus M.S., Superconductivity in bundles of double-wall carbon nanotubes, Sci. Rep. 2 (2012) 625. https://doi.org/10.1038/srep00625.
21. Cao Y., Fatemi V., Fang S., Watanabe K., Taniguchi T., Kaxiras E., Jarillo-Herrero P., Unconventional superconductivity in magic-angle graphene superlattices, Nature 556 (2018) 43–50. https://doi.org/10.1038/nature26160.
22. Benedict L. X., Crespi V. H., Louie S. G., Cohen M. L.. Static conductivity and superconductivity of carbon nanotubes: Relations between tubes and sheets, Phys. Rev. B 52, (1995) 14935. DOI: https://doi.org/10.1103/PhysRevB.52.14935
23. Tang Z.K., Zhang L.Y., Wang N., Zhang X.X., Wen G.H., Li G.D., Wang J.N., Chan C.T., Sheng P.,

Superconductivity in 4 angstrom single-walled carbon nanotubes, Science 292 (2001) 2462–2465.

24. Lortz R., Zhang Q., Shi W., Ye J., Qiu C., Wang Z., He H., Sheng P., Qian T., Tang Z.K., Wang N., Zhang X.X., Wang J., Chan C.T., Superconductivity in ultrathin carbon nanotubes, Proc. Natl. Acad. Sci. U.S.A. 106 (2009) 7299–7303.
25. Ieong C., Wang Z., Shi W., Wang Y., Wang N., Tang Z.K., Sheng P., Lortz R., Observation of the Meissner state in superconducting arrays of 4-Å carbon nanotubes, Phys. Rev. B 83 (2011) 184512. https://doi.org/10.1103/PhysRevB.83.184512.
26. Wang Z., Shi W., Xie H., Zhang T., Wang N., Tang Z., Zhang X., Lortz R., Sheng P., Sheikin I., Demuer A., Superconductivity with extremely high upper critical fields in ultrathin carbon nanotubes, Phys. Rev. B 81 (2010) 174530. https://doi.org/10.1103/PhysRevB.81.174530.
27. Zhang B., Liu Y., Chen Q., Lai Z., Sheng P., Observation of high $T_c$ one dimensional superconductivity in 4 angstrom carbon nanotube arrays, AIP Adv. 7 (2017) 025305. https://doi.org/10.1063/1.4976847.
28. Pan J., Zhang B., Hou Y., Zhang T., Deng X., Wang Y., Wang N., Sheng P., Superconductivity in boron-doped carbon nanotube networks, arXiv:2303.15980 (2023). https://doi.org/10.48550/arXiv.2303.15980.
29. Kokotailo G., Lawton S., Olson D., Meier W.M., Structure of synthetic zeolite ZSM-5, Nature 272 (1978) 437–438. https://doi.org/10.1038/272437a0.
30. Supplementary Materials: Figures S1–S10.
31. Lortz R., Lin F., Musolino N., Wang Y., Junod A., Rosenstein B., Toyota N., Specific heat and magnetic phase diagram of a strongly fluctuating superconductor, Phys. Rev. B 74 (2006) 104502. https://doi.org/10.1103/PhysRevB.74.104502.
32. Pasler V., Schweiss P., Meingast C., Obst B., Wühl H., Rykov A.I., Tajima S., Anomalous thermal expansion and critical fluctuations at the superconducting transition of $YBa_2Cu_3O_{7-\delta}$, Phys. Rev. Lett. 81 (1998) 1094–1097. https://doi.org/10.1103/PhysRevLett.81.1094.
33. Meingast C., Pasler V., Nagel P., Rykov A., Tajima S., Olsson P., Thermodynamic evidence for a vortex-lattice melting transition in $YBa_2Cu_3O_{7-\delta}$, Phys. Rev. Lett. 86 (2001) 1606–1609. https://doi.org/10.1103/PhysRevLett.86.1606.
34. Lortz R., Meingast C., Rykov A.I., Tajima S., Magnetic-Field-Induced Finite-Size Effect in the High-Temperature Superconductor $YBa_2Cu_3O_{7-\delta}$: A Comparison with Rotating Superfluid $^4$He, Phys. Rev. Lett. 91 (2003) 207001. https://doi.org/10.1103/PhysRevLett.91.207001.
35. 34. Schilling A., Fisher R.A., Phillips N.E., Welp U., Kwok W.K., Crabtree G.W., Calorimetric determination of the magnetic-field–temperature phase diagram of $YBa_2Cu_3O_{7-\delta}$, Phys. Rev. Lett. 78 (1997) 4833–4836. https://doi.org/10.1103/PhysRevLett.78.4833.
36. Blonder G.E., Tinkham M., Klapwijk T.M., Transition from metallic to tunneling regimes in superconducting microconstrictions: excess current, charge imbalance, and supercurrent conversion, Phys. Rev. B 25 (1982) 4515–4532. https://doi.org/10.1103/PhysRevB.25.4515.
37. Wang Y., Plackowski T., Junod A., Specific heat in the superconducting and normal state (2–300 K, 0–16 T), and magnetic susceptibility of the 38 K superconductor MgB2: evidence for a multicomponent gap, Physica C 355 (2001) 179–193. https://doi.org/10.1016/S0921-4534(01)00617-7.
38. Kostić P., Veal B., Paulikas A.P., Welp U., Paramagnetic Meissner effect in Nb, Phys. Rev. B 53 (1996) 791–795. https://doi.org/10.1103/PhysRevB.53.791.
39. Pearl J., Current distribution in superconducting films carrying quantized fluxoids, Appl. Phys. Lett. 5 (1964) 65–66. https://doi.org/10.1063/1.1754056.
40. Nguyen A.K., Sudbø A., Topological phase fluctuations, amplitude fluctuations, and criticality in extreme type-II superconductors, Phys. Rev. B 60 (1999) 15307–15327. https://doi.org/10.1103/PhysRevB.60.15307.
41. Tešanović Z., Charge-vortex duality and the phase diagram of cuprate superconductors, Phys. Rev. B 59 (1999) 6449–6474. https://doi.org/10.1103/PhysRevB.59.6449.
42. Wang Y., Li L., Ong N.P., Nernst effect in high-Tc superconductors, Phys. Rev. B 73 (2006) 024510. https://doi.org/10.1103/PhysRevB.73.024510.
43. Eagles D.M., Possible pairing without superconductivity at low carrier concentrations in bulk and thin-film superconducting semiconductors, Phys. Rev. 186 (1969) 456–463. https://doi.org/10.1103/PhysRev.186.456.
44. Božović I., Levy J., Pre-formed Cooper pairs in copper-oxide superconductors, Nat. Phys. 16 (2020) 712–717. https://doi.org/10.1038/s41567-020-0915-8.

45. Daghero D., Gonnelli R.S., Probing multiband superconductivity by point-contact spectroscopy, Supercond. Sci. Technol. 23 (2010) 043001. https://doi.org/10.1088/0953-2048/23/4/043001.
46. Meyer B., Elsaesser C., Faehnle M., FORTRAN90 program for mixed-basis pseudopotential calculations for crystals, Max-Planck-Institut für Metallforschung, Stuttgart, unpublished.
47. Heid R., Bohnen K.P., Linear response in a density-functional mixed-basis approach, Phys. Rev. B 60 (1999) R3709–R3712. https://doi.org/10.1103/PhysRevB.60.R3709.

**Author Contributions**

PS investigated the superconductivity of carbon nanotubes for the past 25 years; he proposed an in situ boron doping approach, provided guidance on the fabrication of the samples that led to the present results, and coordinated the research effort. NW and PS initiated the present investigation, followed by RL, who provided the experimental expertise to detect high-temperature superconductivity in this material. YW, RH, and WMC fabricated the samples. YH and JP implemented the initial boron doping setup. NW improved upon the boron-doping setup and procedure. YW and RL performed the DC magnetization measurements. YW carried out sample characterization. NW proposed the 2-probe macroscopic measurements of electrical transport, and the approach was refined by YW and RL in the form of the 4-probe set-up. RL, THK, YHN and YW performed the transport measurements. RL and THK investigated the pressure dependence; TTL, YHN and RL performed the specific heat measurements; and THK and RL performed the point contact spectroscopy measurements. SK and RL prepared the setup for the simultaneous measurements of the AC susceptibility and electrical resistivity. THK and RL performed the AC susceptibility measurements with the help of A. Demuer. RL and THK analyzed all the experimental data, providing evidence for superconductivity. TZ performed the ab initio calculations. PS interpreted the Raman spectra on the basis of the results of the ab initio calculations. RL wrote the manuscript, with contributions from PS and YW. RL and PS penned the final draft.

# SUPPLEMENTARY MATERIALS

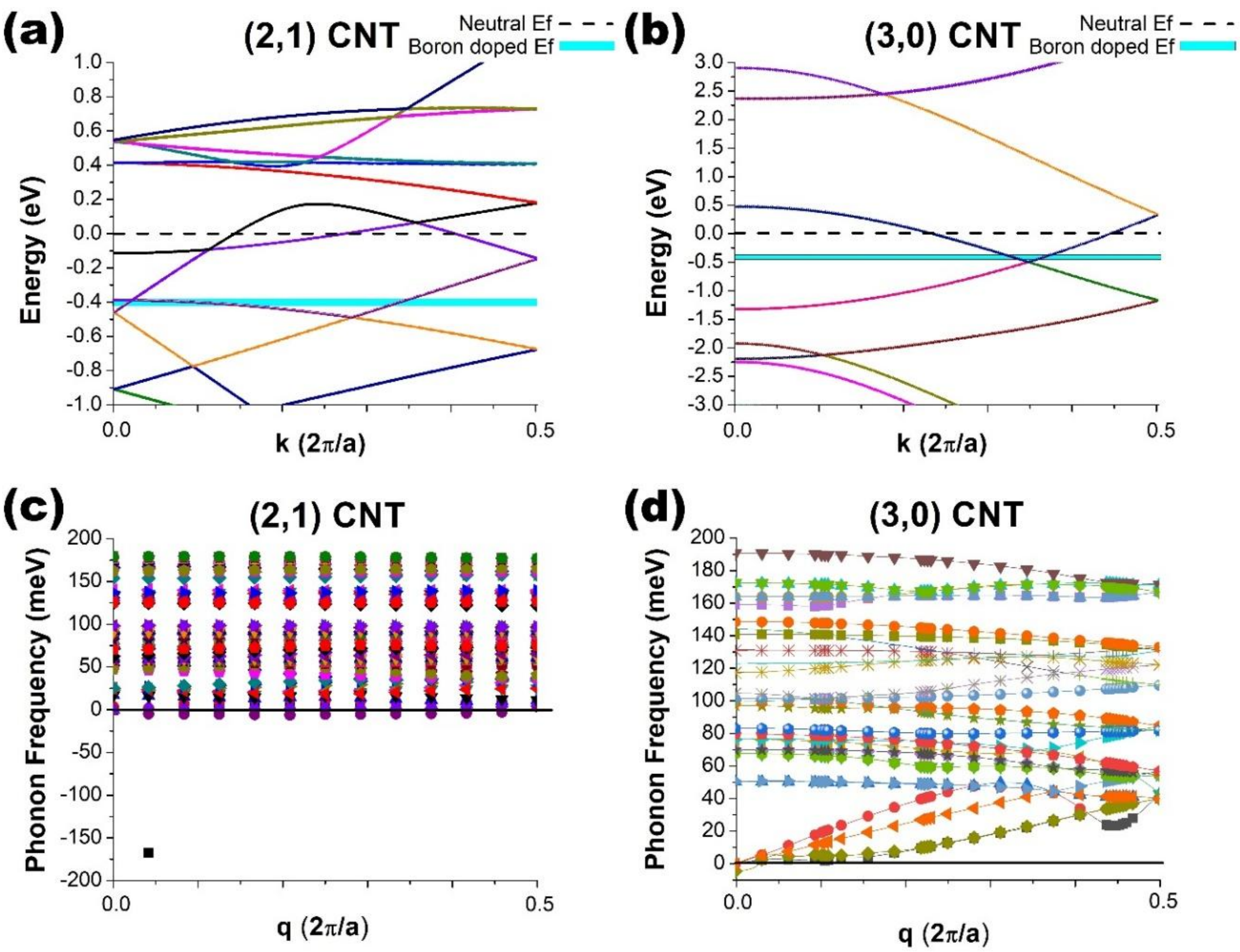

**Fig. S1. Electronic band structure and phonon spectrum calculations.**
**a,** Electronic band structure of the (2, 1) CNT. The black solid line denotes the Fermi level in the neutral (undoped) state. The shaded blue region represents the range of Fermi levels observed under boron doping, which is consistent with the experimental doping ratios. Upon doping, the Fermi level shifts toward the van Hove singularity, which is located at a flat band near -4.78 meV at $k = 0$. **b,** Electronic band structure of the (3, 0) CNT. Both the (2, 1) and (3, 0) CNTs exhibit metallic behavior, as evidenced by the partially filled bands crossing the Fermi level. **c,** Calculated phonon spectrum of the undoped (2, 1) CNT. The zero-frequency line is shown as a solid black line for reference. The imaginary phonon modes (plotted as negative frequencies) indicate the dynamic instability of the isolated (2, 1) CNT, suggesting that structural stability is maintained only when the CNTs are confined by the channel wall. Owing to the presence of imaginary modes, no q-point interpolation was applied. **d,** Phonon spectrum of the undoped (3, 0) CNT. The softening of the phonon modes at specific nesting wavevectors is observed, indicating electron–phonon coupling effects.

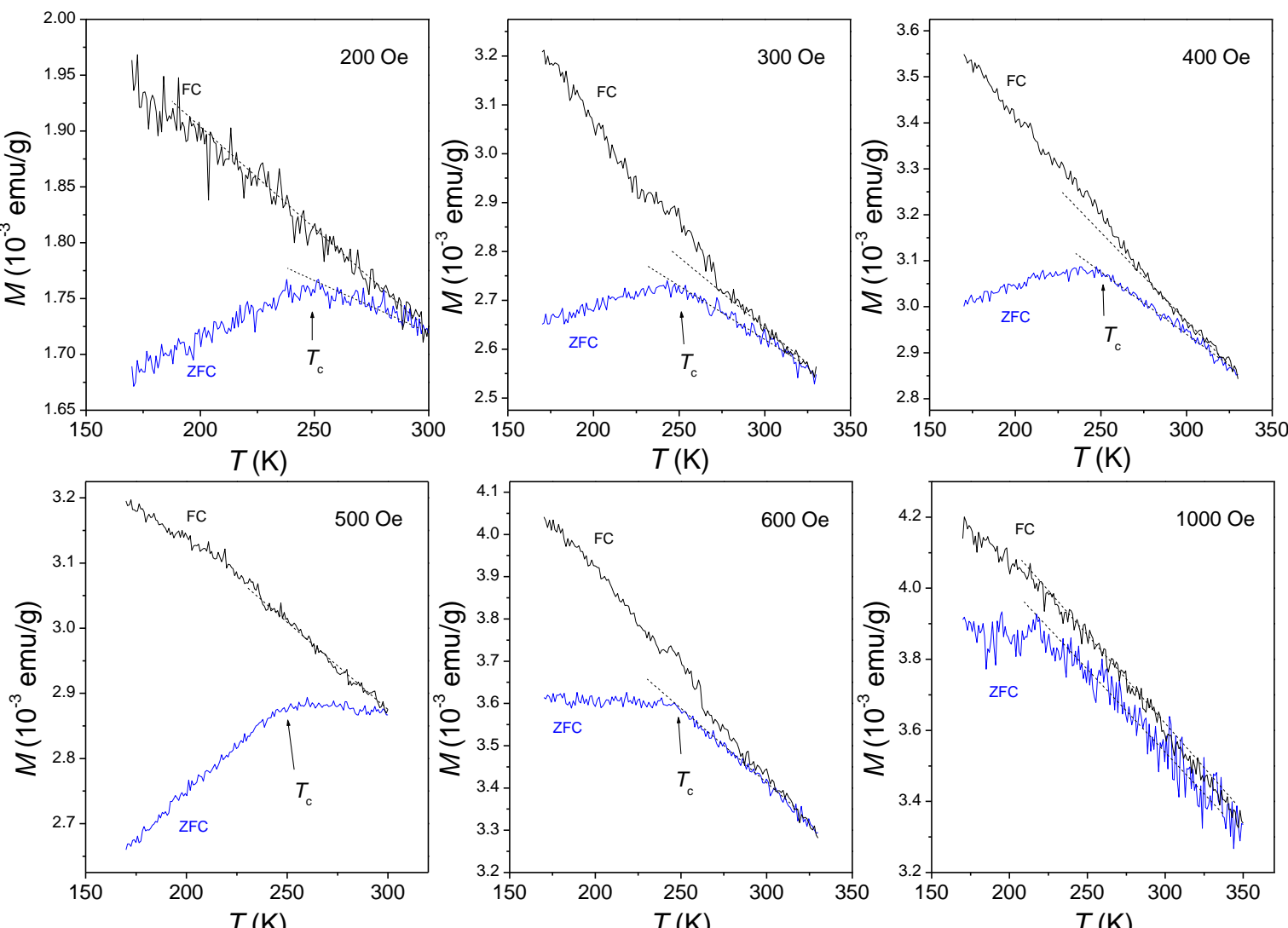

**Fig. S2. Selected raw data of the DC magnetization of batch #61 before removal of the linear background.**

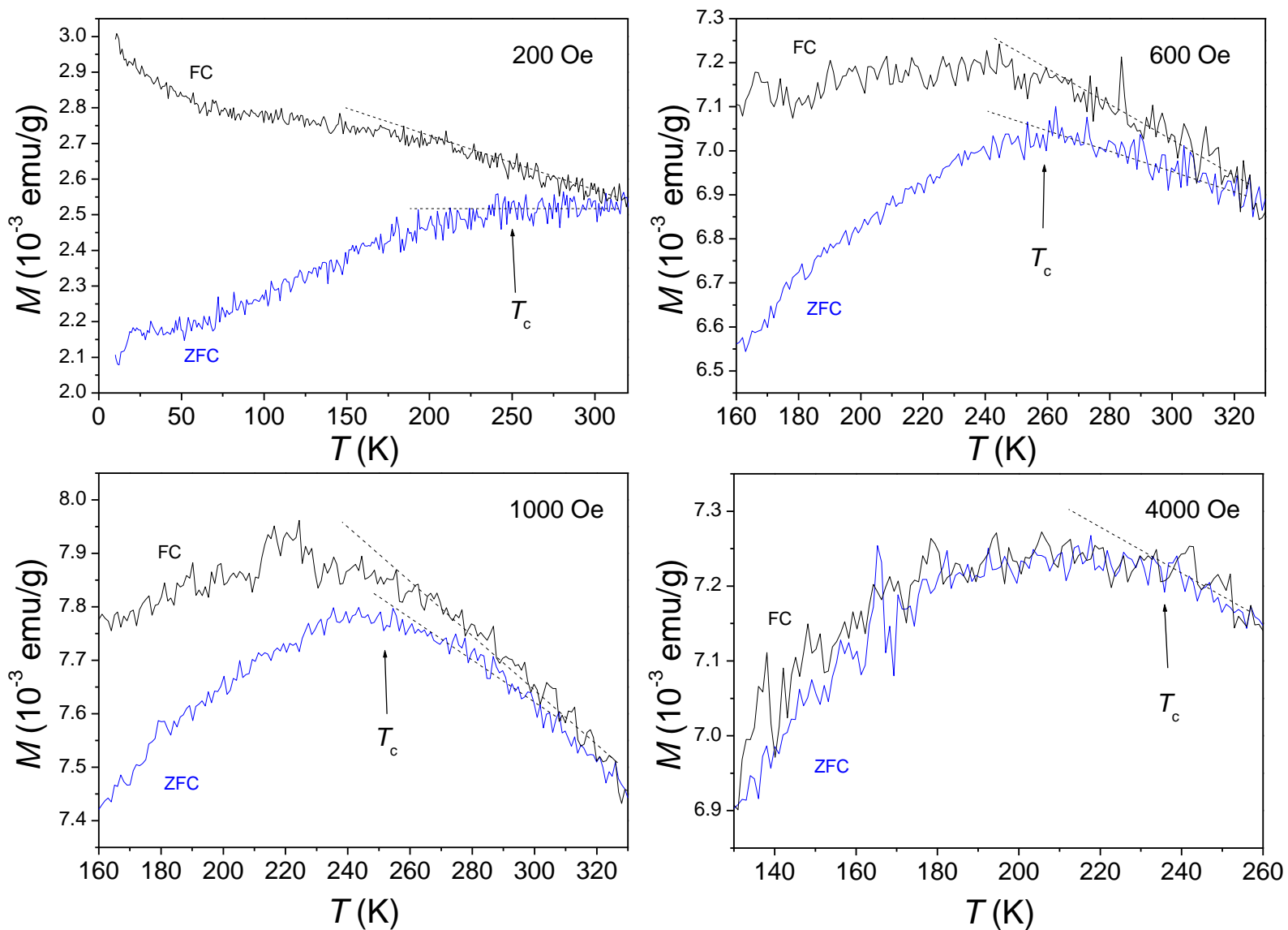

**Fig. S3. Selected raw data of the DC magnetization of batch #62 before removal of the linear background.**

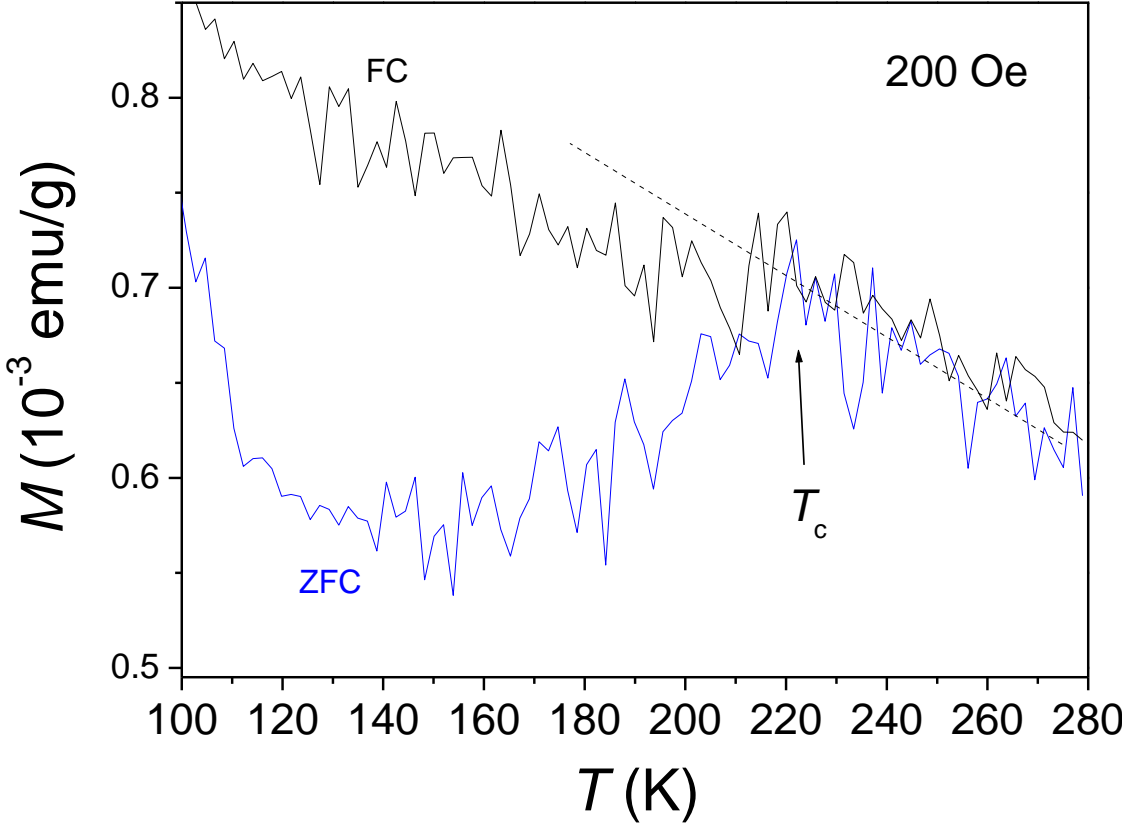

**Fig. S4. Raw data of the 200 Oe DC magnetization of batch #66 before removal of the linear background.**

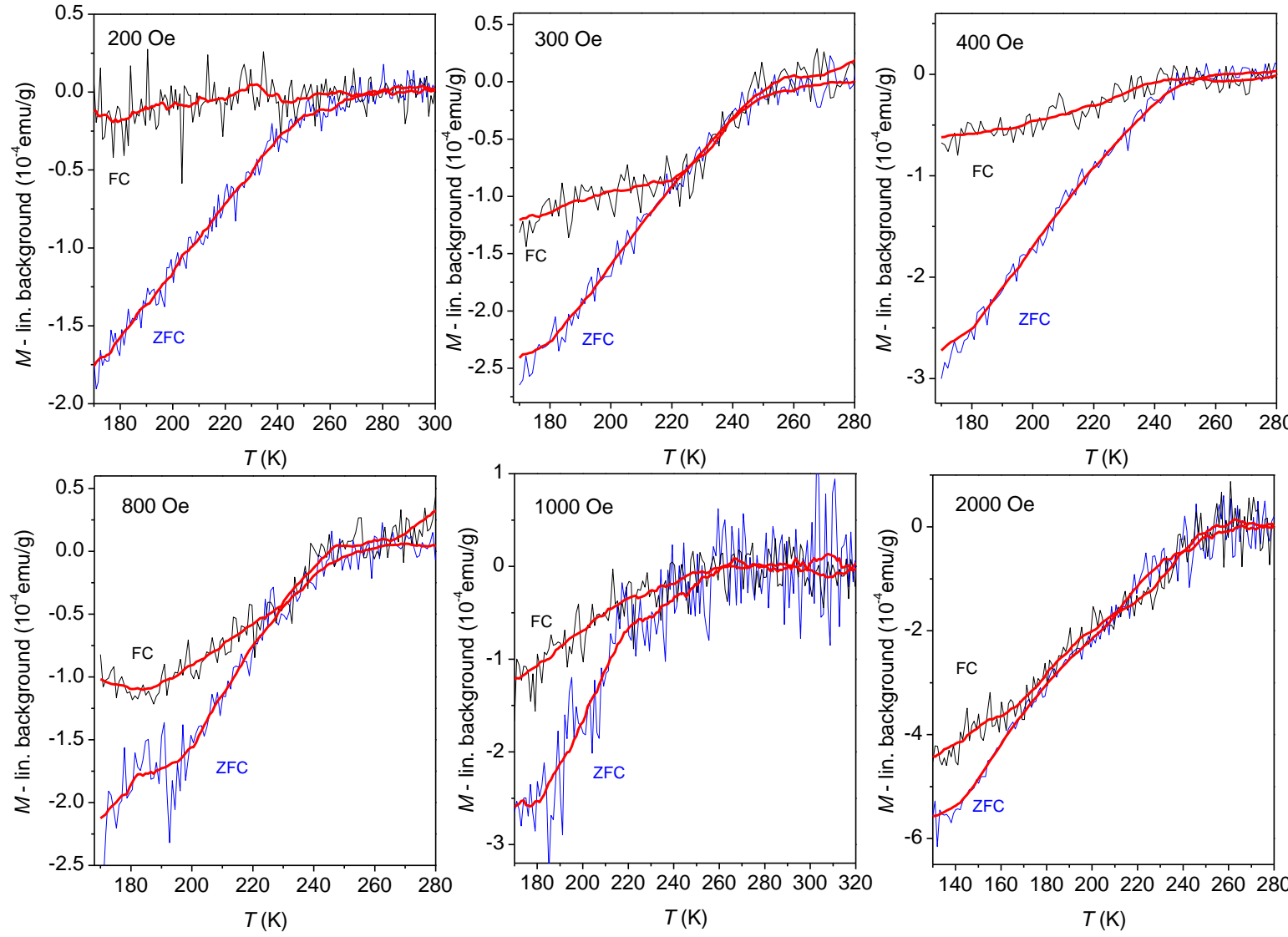

**Fig. S5. Supplementary DC magnetization data of batch #61.**

The data were acquired under ZFC (blue) and FC (black) conditions with various applied fields. The red lines indicate smoothed data obtained by adjacent averaging over 5–10 data points.

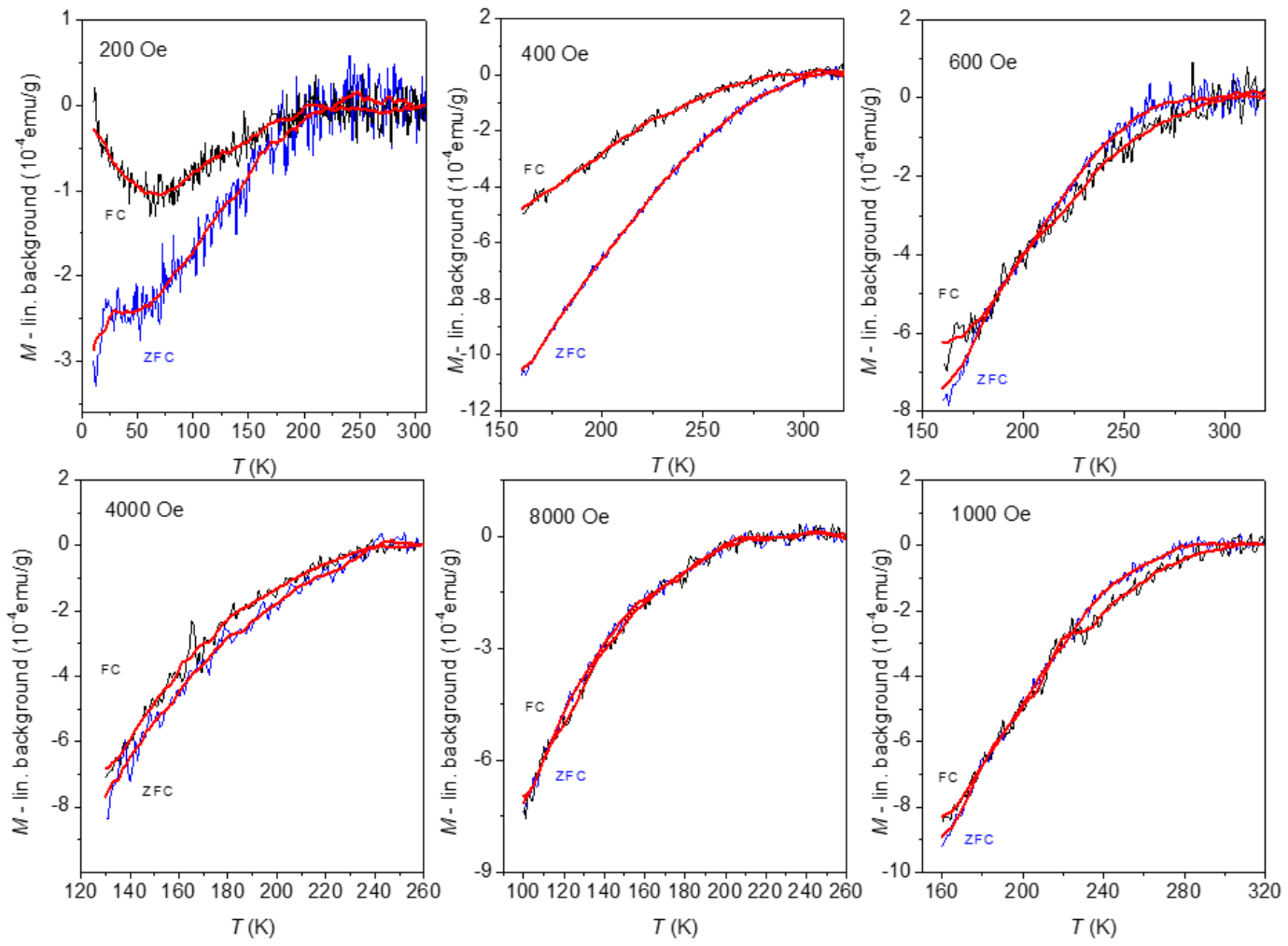

**Fig. S6. Supplementary DC magnetization data of batch #62.**
The data were acquired under ZFC (blue) and FC (black) conditions with various applied fields. The red lines represent smoothed data obtained by adjacent averaging over 5–10 data points.

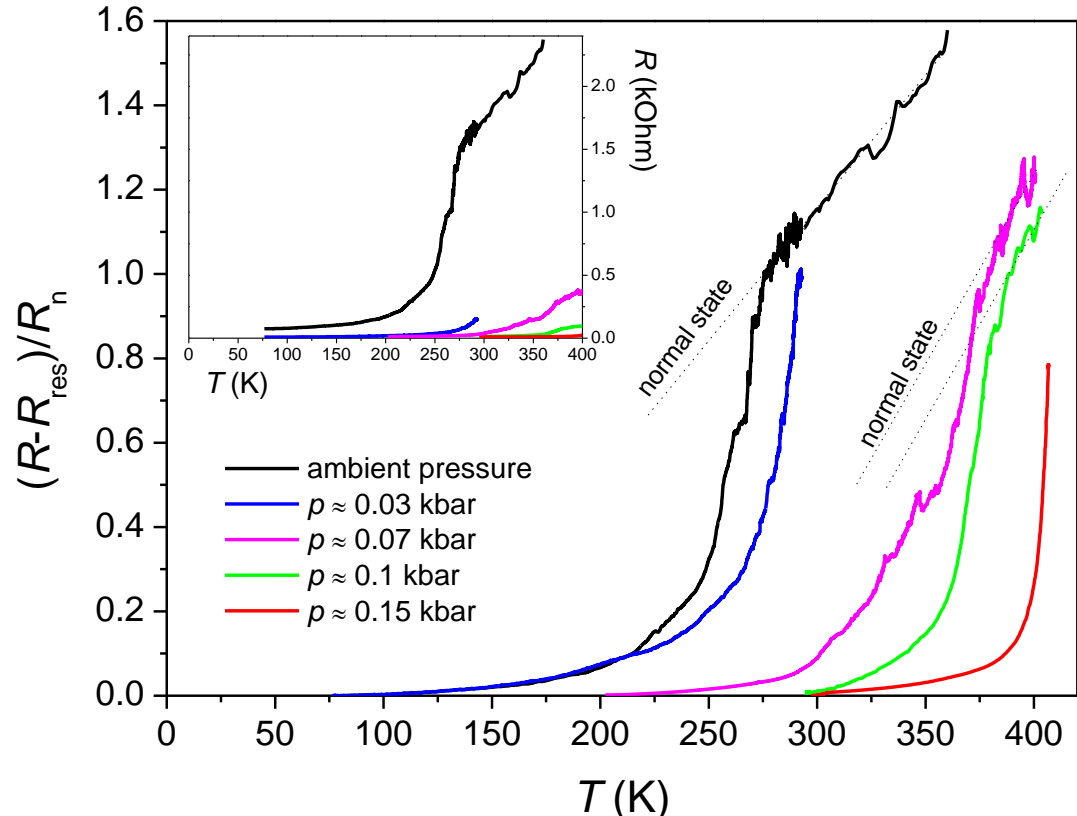

**Fig. S7. Supplementary 2-probe pressure dependent resistance data.**
Plot of the normalized electric resistivity of batch #66 measured with the 2-probe device under various pressures. Note that the pressure values are indicative and estimated from the torque applied to the screws. The normalization of the 0.15 kbar data is approximate, as we were unable to reach the normal state above 400 K with our setup.

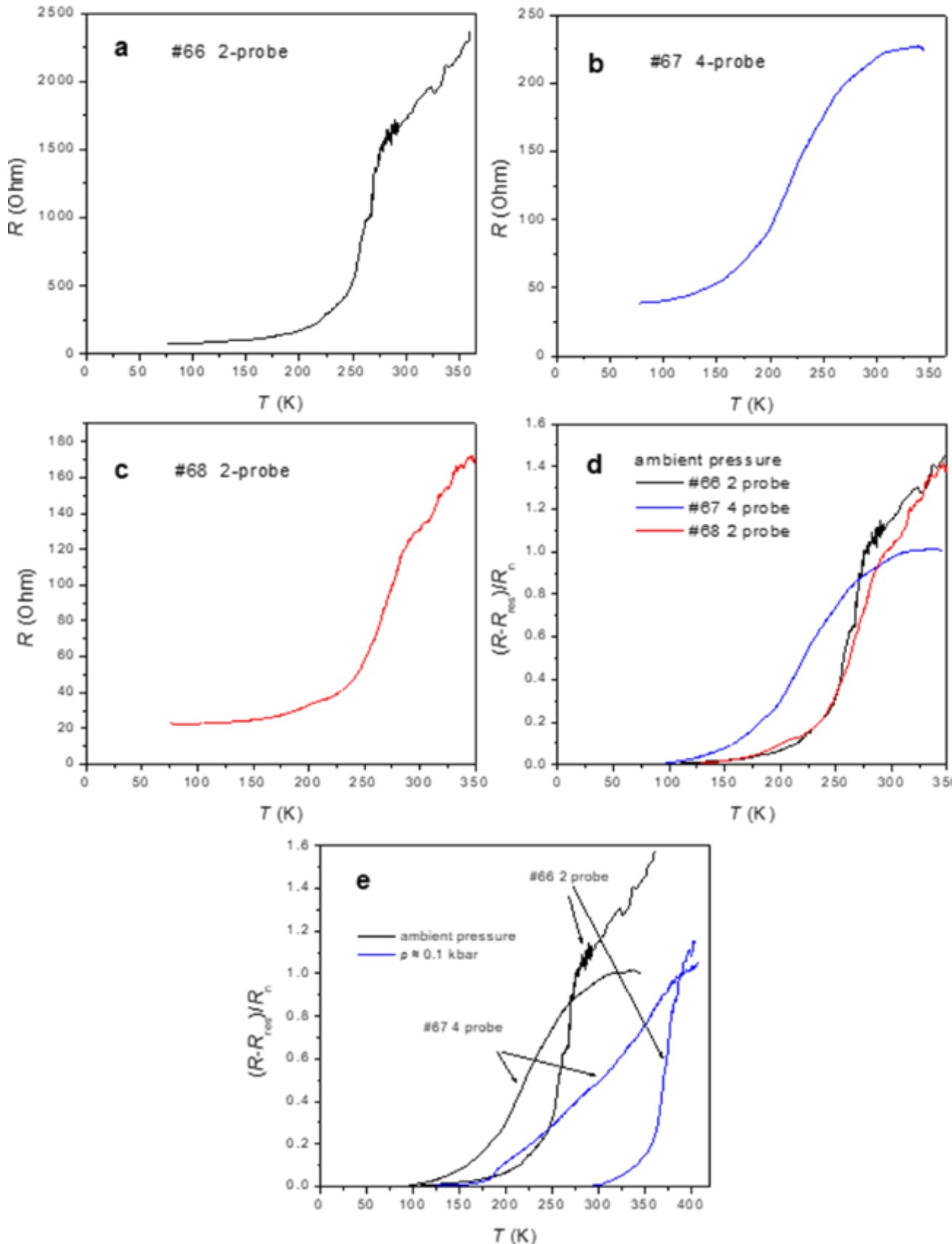

**Fig. S8. Supplementary resistance data.**
Raw ambient pressure resistance data of batches #67 (**a**, 2 probes), #68 (**b**, 4 probes) and #68 (**c**, 2 probes). **d** Comparison of the normalized ambient pressure resistance data with the residual resistance removed for batches #67, #67 and #68. **e** Normalized resistance data with the residual resistance removed for batch #67, measured with the 4-probe device at zero pressure and ~0.1 kbar; used for comparison with the 2-probe data of batch #66 shown in the main article. Note that the different widths of the ambient pressure transitions are likely caused by disorder in batch #67 and not by the 4-probe technique. The large width of the #67 pressure data indicates a large pressure gradient owing to the use of the 4-probe device, making it unsuitable for pressure experiments.

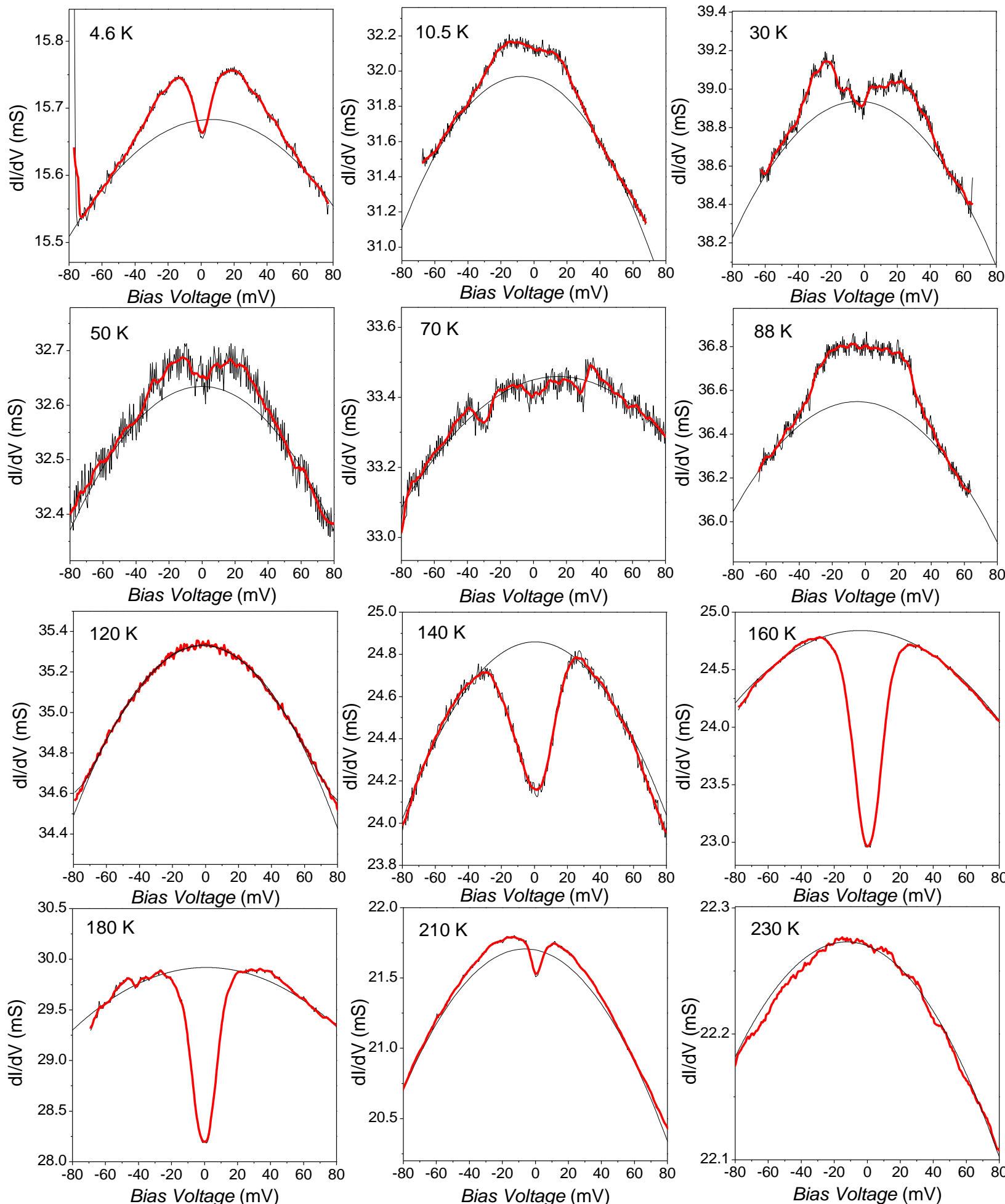

**Fig. S9. Selection of raw point–contact data obtained at various temperatures with fits of the parabolic background contribution.** The black lines represent the raw data; the red lines represent the data smoothed by adjacent averaging over 5–10 data points.

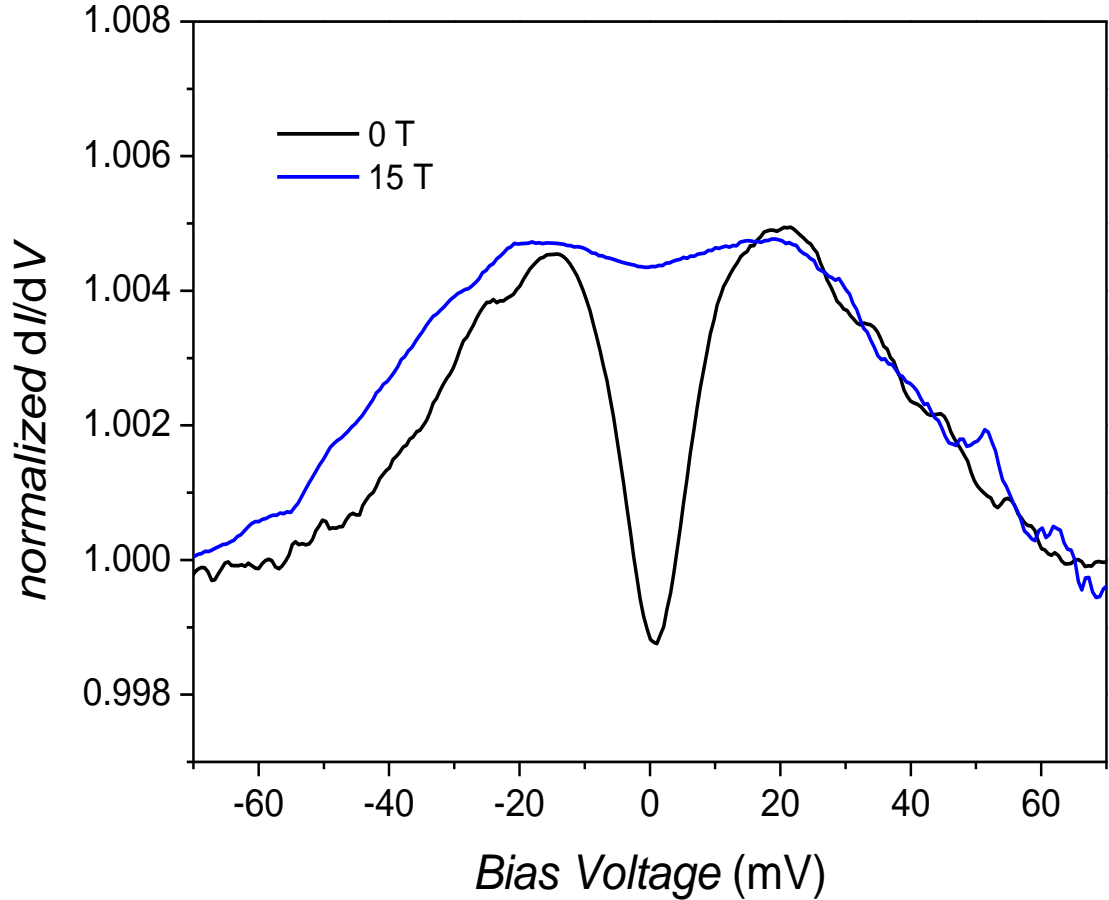

**Fig. S10. Magnetic field dependence of the point–contact spectra of CNT@ZSM-5 at 4.6 K.** Within the limits of experimental uncertainty, no variation in the superconducting gap is observed, suggesting an exceptionally high upper critical field, likely exceeding 100 T. The apparent asymmetry in the 0 T data is an experimental artifact attributed to minor changes in the tip position.